\documentclass[jcp,amsmath,amssymb,showkeys,
reprint,
]{revtex4-2}
\usepackage[utf8]{inputenc}

\usepackage{bibentry}
\usepackage[toc,page]{appendix}
\usepackage[nonumberlist, acronym, toc]{glossaries}

\usepackage[caption=false]{subfig}
\usepackage{tabularx}
\usepackage{algpseudocode, algorithm}

\usepackage{physics}
\usepackage{mathtools}
\usepackage{bm}

\usepackage{enumerate}

\newcommand{\mb}[1]{\mathbf{#1}}

\newcommand{\pd}{\partial}
\newcommand{\schro}{Schr{\"o}dinger }
\DeclareMathOperator{\EX}{\mathbb{E}}

\usepackage{stackengine}

\usepackage{xcolor}

\DeclareFontFamily{U}{solomos}{}
\DeclareErrorFont{U}{solomos}{m}{n}{10}
\DeclareFontShape{U}{solomos}{m}{n}{
  <-> s*[1.1]  gsolomos8r
}{}
\newcommand{\vkappa}{\text{\usefont{U}{solomos}{m}{n}\symbol{'153}}}

\newacronym{kfac}{KFAC}{Kronecker-Factored Approximate Curvature}
\newacronym{qmc}{QMC}{Quantum Monte Carlo}
\newacronym{dmc}{DMC}{Diffusion Monte Carlo}
\newacronym{gpu}{GPU}{Graphics Processing Unit}
\newacronym{vmc}{VMC}{Variational Monte Carlo}
\newacronym{mcmc}{MCMC}{Markov Chain Monte Carlo}

\newacronym{bm}{BM}{Boltzmann Machine}
\newacronym{vae}{VAE}{Variational Autoencoder}
\newacronym{sr}{SR}{Stochastic Reconfiguration}

\newacronym{hf}{HF}{Hartree Fock}

\newacronym{mps}{MPS}{Matrix Product States}

\newacronym{fim}{FIM}{Fisher Information Matrix}
\newacronym{iad}{IAD}{Independent Activations and Sensitivities}

\begin{document}

\title{Simulations of state-of-the-art fermionic neural network wave functions with diffusion Monte Carlo}
\author{Max Wilson}
\email{aw16952@bristol.ac.uk}
\affiliation{Quantum Engineering CDT, Bristol University, Bristol, BS8 1TH, UK}

\affiliation{QuAIL, NASA Ames Research Center, Moffett Field, California 94035, USA}

\affiliation{KBR, Inc., 601 Jefferson St., Houston, TX 77002, USA}

\author{Nicholas Gao} 
\affiliation{QuAIL, NASA Ames Research Center, Moffett Field, California 94035, USA}

\affiliation{KBR, Inc., 601 Jefferson St., Houston, TX 77002, USA}

\affiliation{German Aerospace Center (DLR),
Linder H\"ohe
51147 K\"oln, Germany
}
\affiliation{Technical University of Munich,
Boltzmann Str. 3
85748 Garching, Germany}

\author{Filip Wudarski}
\email{filip.a.wudarski@nasa.gov}

\affiliation{QuAIL, NASA Ames Research Center, Moffett Field, California 94035, USA}
\affiliation{USRA Research Institute for Advanced Computer Science, Mountain View, California 94043, USA}

\author{Eleanor Rieffel}
\affiliation{QuAIL, NASA Ames Research Center, Moffett Field, California 94035, USA}

\author{Norm M. Tubman}
\email{norm.m.tubman@nasa.gov}
\affiliation{QuAIL, NASA Ames Research Center, Moffett Field, California 94035, USA}

\date{\today}
\begin{abstract}
    Recently developed neural network-based \emph{ab-initio} solutions (Pfau et. al arxiv:1909.02487v2) for finding ground states of fermionic systems can generate state-of-the-art results on a broad class of systems. In this work, we improve the results for this Ansatz with Diffusion Monte Carlo. Additionally, we introduce several modifications to the network (Fermi Net) and optimization method (Kronecker Factored Approximate Curvature) that reduce the number of required resources while maintaining or improving the modelling performance. In terms of the model, we remove redundant computations and alter the way data is handled in the permutation equivariant function.  The Diffusion Monte Carlo results exceed or match state-of-the-art performance for all systems investigated: atomic systems Be-Ne, and the carbon cation C$^+$.  
\end{abstract}
\maketitle



\section{Introduction}
Neural networks, in recent years, have provided  an alternative computational paradigm for solving electronic structure problems which includes applications in directly solving the time-independent \schro equation to  find approximate ground states of atoms and molecules \cite{von2020retrospective}. Standard quantum chemistry methods \cite{szabo2012modern}, such as coupled-cluster \cite{kummel2003biography}, full configuration interaction \cite{ross1952calculations}, \gls{vmc} \cite{mcmillan1965ground}, and \gls{dmc} \cite{umrigar1993diffusion} have been used in conjunction with neural network methods in different ways, such as the introduction of new Ans{\"a}tze \cite{pfau2019ab, troyer2005computational, hermann2019deep, schutt2019unifying, schutt2017schnet, carleo2017solving} or providing rich datasets for the prediction of properties of previously untested systems and their dynamics in supervised learning frameworks \cite{balabin2009neural, goh2017deep, cova2019deep, yang2020machine}.

Notable examples of new Ans{\"a}tze are neural networks such as SchNet \cite{schutt2017schnet}, PauliNet \cite{hermann2019deep}, Boltzmann machines \cite{carleo2017solving} and Fermi Net \cite{pfau2019ab}. Even though these techniques are still in the early stages of development, machine learning for fermionic systems has been widely studied over the past few years, and some of these techniques are capable of producing state-of-the-art results for electronic structure simulations. We have particular interest in finding good approximations to ground states of fermionic Hamiltonians \cite{hermann2019deep} and the precision and accuracy of Fermi Net \cite{pfau2019ab}. 

In this paper we modify an existing framework, the Fermi Net \cite{pfau2019ab}, by changing how data is handled in the network and removing redundant elements.  We apply the wave function optimization algorithm \gls{vmc} whilst altering some aspects of the \gls{kfac} optimization, and then running the \gls{dmc} algorithm to improve the wave function further. The use of \gls{dmc} is standard practice in many \gls{qmc} codes, however, the wave functions used in virtually all packages prior to this new wave of neural network approaches are using a wave function consisting of Slater/Jastrow/Multi-determinant/Backflow components, which generally are less accurate or less systematically improveable than the Fermi Net Ansatz.   

\begin{figure}
    \centering
    \includegraphics[width=0.4\textwidth]{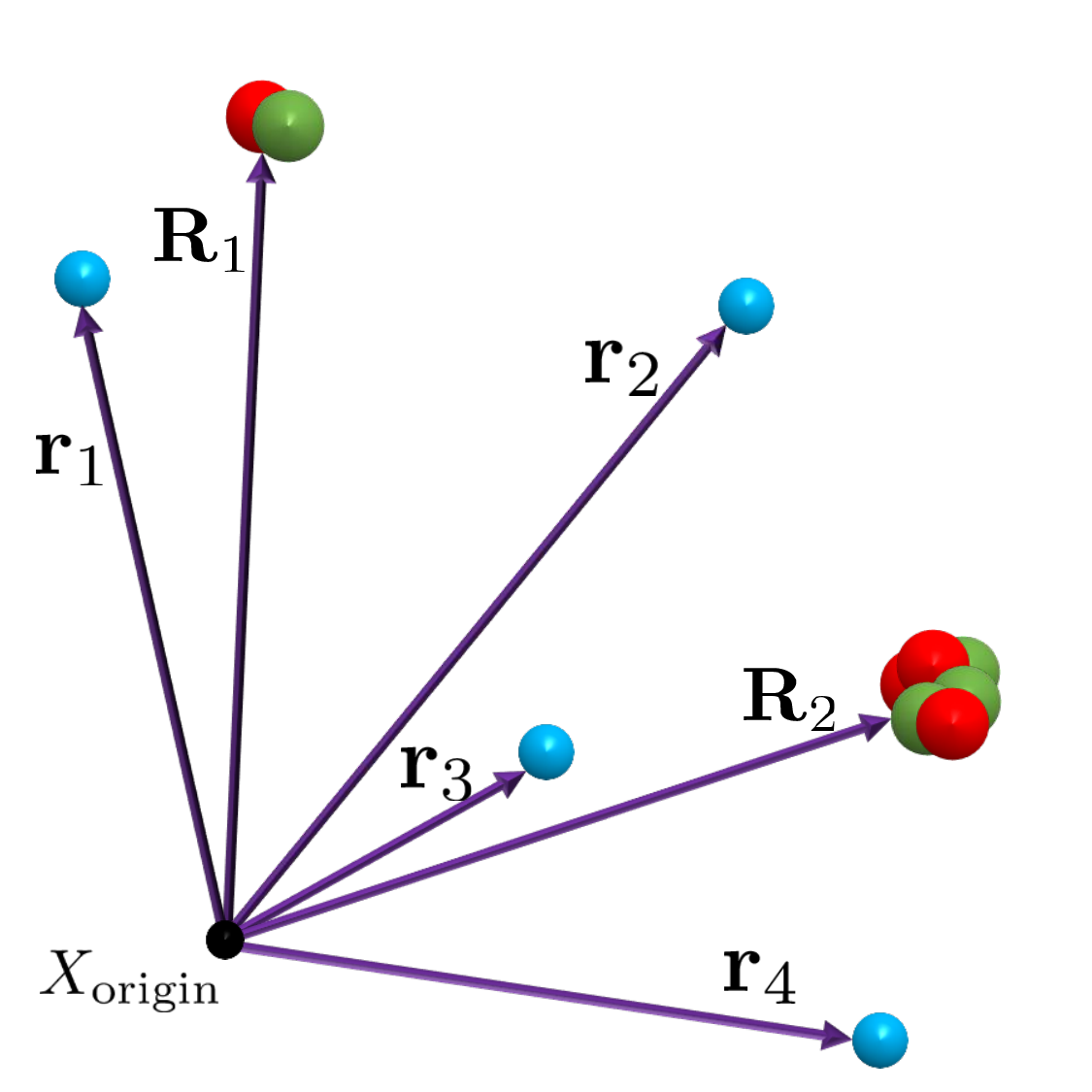}
    \caption{The system of atoms (protons/neutrons red/green) and electrons (blue). The set of position vectors of a system of atoms, $\mb{R}_i$ and electrons $\mb{r}_i$. $X_\text{origin}$ is the origin  (black) of the coordinate system. In all cases explored here we considered single atom systems and the origin was set to the nucleus position. However, the choice of origin is completely arbitrary as Fermi Net is invariant to translations. 
    }
    \label{fig:system}
\end{figure}

The \gls{qmc} we described above is generally run with standard \gls{vmc} and then \gls{dmc} with the fixed node approximation (nuclei are fixed in place) in the continuum.  That is to say we solve the \schro equation 
\begin{equation}
    \hat{H} \psi(X) = E \psi(X), 
\end{equation}
where $X$ defines the system configuration, Figure~\ref{fig:system}, $\hat{H}$ is the Hamiltonian operator and $E$ is the energy of the eigenfunction $\psi(X)$, the wave function we are attempting to model. The Ansatz developed in this work is functionally identical to previous work, although improvements are made. We refer to this model and the associated optimization methods as Fermi Net* in order to distinguish from previous work referred to as Fermi Net.

The paper is organized as follows: Section~\ref{sec:schrodinger} introduces the problem to be solved (i.e. the time-independent \schro equation for fermionic systems) and all the relevant background including \gls{vmc}, \gls{kfac}, \gls{dmc} and a sketch of the Fermi Net;  Section~\ref{sec:methods} describes the particular methods used in this work including detailed descriptions of the algorithms; The results are described and discussed in Section~\ref{sec:results}; and finally we conclude our findings in Section~\ref{sec:conclusions}.

There are several contributions in this work:
\begin{itemize}
    \item Introduced changes to the Fermi Net implementation: removing the diagonal elements of the pairwise terms; and changing how the data is handled in the permutation equivariant function. Both changes result in efficiency improvements;
    \item first (to our knowledge) application of diffusion Monte Carlo with a neural network Ansatz;
    \item and state-of-the-art results on all systems explored (Be-Ne, C$^+$)
\end{itemize}

\section{Solving the \schro equation for fermionic systems}\label{sec:schrodinger}

\subsection{Variational Monte Carlo}

The Schr\"odinger Equation plays a central role in the description of quantum behavior of chemical systems. 
Apart from a handful of analytically solvable models, for example the hydrogen atom \cite{bransdenquantum}, one mostly needs to incorporate approximate techniques and numerical methods that scale unfavorably with the increasing system size (e.g. number of electrons) \cite{troyer2005computational}.  Many techniques and approximations have been introduced to address this problem, and in this work we have a particular focus on real space Monte Carlo approaches.

An atomic or molecular system can be described by the time-independent \schro  equation
\begin{equation}\label{eq:schrodinger}
    \hat{H} \psi(X) = E \psi(X).
\end{equation}
Using the Born-Oppenheimer approximation, the position of nuclei are frozen and we have a Hamiltonian for the electronic degrees of freedom,
\begin{equation}
    \hat{H} = -\frac{1}{2} \hat{\mb{\nabla}}^2 + V(X).
\end{equation}
Units of energy $E$ expressed here are Hartrees, $\hat{\mb{\nabla}}^2$ is the multidimensional ($3n_e$ where $n_e$ is the number of electrons) Laplacian of the wave function $\hat{\mb{\nabla}}^2 = \sum_{i=1}^{n_e}\sum_{j=x,y,z} \frac{d^2}{dr_{i,j}^2}$ that describes kinetic energy. $r_{i,j}$ and $R_{i,j}$ correspond to coordinate $j=x,y,z$ of the $i$-th electron and nuclei, respectively. $V(X)$ is the potential energy of some (electron, nuclei) configuration 
\begin{equation}
X = (r_{1,x},r_{1,y},r_{1,z},\ldots,r_{n_e,z}; R_{1,x},\ldots, R_{n_n,z}),
\end{equation}
given by
\begin{align}
    V(X) =& \sum_{i>j}^{n_e} \frac{1}{|\mb{r}_i - \mb{r}_j|}  + \sum_{i,I}^{n_e,n_n} \frac{Z_I}{|\mb{r}_i - \mb{R}_I|} \nonumber \\ &+ \sum_{I>J}^{n_n} \frac{Z_IZ_J}{|\mb{R}_I - \mb{R}_J|}
\end{align}
where $n_n$ is the number of nuclei, $Z_I$ is the atomic number of nuclei $I$, and $\mb{r}_i$ and $\mb{R}_I$ are the position vectors of the electron $i$ and nuclei $I$, respectively. Throughout the paper we use bold font to denote vectors and regular font to denote vector components (scalars). Both are lower case symbols or letters. Matrices are capitalized symbols or letters and are not written in bold font. Though a batched vector, for example $m$ vectors of dimension $n$ arranged in an $m \times n$, can be represented as a matrix, we notationally treat it as a vector, as is typical for deep learning. There are some caveats to these, though in general the meaning is obvious from context, for example nuclei coordinates are capitalized. 

Approximately solving the \schro equation, Equation~\eqref{eq:schrodinger}, is a subroutine in finding the minimum energy of the system, i.e. the ground state energy $E_0$. Since the Hamiltonian is a bounded operator, one may use a variational principle \cite{szabo2012modern} 
\begin{equation}
    E_0 \le \frac{\bra{\psi}\hat{H}\ket{\psi}}{\braket{\psi}}=\frac{\int d X \psi^*(X)\hat{H}\psi(X)}{\int dX \psi^*(X)\psi(X)},
\end{equation}
detailing that the expectation value of the Hamiltonian $\hat{H}$ with respect to a state $\psi(X)$ is bounded from below by the ground state energy $E_0$. Finding the best approximation to the ground state $\psi_0(X)$ can be done with a parameterized Ansatz, a so-called trial wave function $\psi(X;\theta)$, which is iteratively optimized until a satisfactory accuracy (in terms of energy) is achieved. Different Ans\"atze have varying capacities to express wave functions, resulting in different possible minimal energy wave functions. The greater the capacity of an Ansatz to model the true wave function, the better the approximation to the ground state, in general. 

A popular class of variational methods - Variational Monte Carlo (VMC) - relies on random sampling of the configuration space in order to estimate expectation value of the Hamiltonian (computing loss function) as
\begin{equation}\label{eq:loss_function_vmc}
    \mathcal{L}(\theta) = \frac{\bra{\psi(\theta)}\hat{H}\ket{\psi(\theta)}}{\braket{\psi(\theta)}} = \frac{\int d X |\psi(X;\theta)|^2 E_L(X;\theta)}{\int dX |\psi(X;\theta)|^2},
\end{equation}
where $E_L(X;\theta) = \psi^{-1}(X;\theta)\hat{H} \psi(X;\theta) $ is local energy, which for molecular/atomic Hamiltonians is convenient to express in log-domain as
\begin{align}
    E_L(X';\theta) = -&\frac{1}{2} \Big[ \hat{\mb{\nabla}}^2 \log|\psi(X;\theta)|\big|_{X'} \nonumber \\ +& \big(\hat{\mb{\nabla}} \log|\psi(X;\theta)|\big|_{X'}\big)^2 \Big] + V(X').
\end{align}
$(\hat{\nabla} \cdot)^2$ is the inner product of the nabla operator $\Big(\frac{\partial \cdot}{\partial r_{1,x}}, ..., \frac{\partial \cdot}{\partial r_{n_e,z}}\Big)$ with itself. 

The integral \eqref{eq:loss_function_vmc} is an expectation value of the sampled configurations $X$, 
\begin{eqnarray}
    \int dX |\psi(X;\theta)|^2 E_L(X;\theta) = \EX_{X\sim p(X;\theta)}\Big[E_L(X;\theta)\Big].
    \label{eq:int_expectation}
\end{eqnarray}
The expectation in Equation~\eqref{eq:int_expectation} is approximated by a Monte-Carlo estimate, 
\begin{align}
    \EX_{X\sim p(X;\theta)}\Big[E_L(X;\theta)\Big]
    \approx \frac{1}{N} \sum_{i=1}^N E_L(X_i;\theta),
\end{align}
where we introduce configuration probability $p(X;\theta)\propto|\psi(X;\theta)|^2$. Samples (also referred to as walkers and are represented by $X$) are generated from the wave function distribution via the Metropolis Hastings Monte Carlo method. 
In order to update the parameters $\theta$ and improve the wave function, one needs to compute gradients of the loss function with respect to $\theta$ denoted $\Delta\mathcal{L}(\theta)$.
The parameters $\theta$ of the wave function are optimized using some form of gradient descent and computed via
{\footnotesize 
\begin{align}
    \Delta \mathcal{L}(\theta) = \EX_X \Big[(E_L(X;\theta) - \EX_X[E_L(X;\theta)])\hat{\mb{\nabla}} \log|\psi(X;\theta)|  \Big] \label{eq:vmc_gradients}
\end{align}}
and estimated through sampling of the configuration space. This procedure allows us to get close to the ground state $\psi_0(X)$, however it strongly relies on the parameterized Ansatz and ease of computing the gradients $\Delta \mathcal{L}(\theta)$. From now on, we will omit $\theta$ parameters where it is clear from the context, and introduce Fermi Net as an Ansatz that provides powerful parameterization.

\subsection{Fermionic Neural Network Ansatz}\label{sec:ferminet}

The Fermionic Neural Network (Fermi Net) is a neural network designed specifically for the task of representing the wave function, in continuous Euclidean space, of the \schro equation for a fermionic Hamiltonian. In this section we describe the original model, found in Reference~\cite{pfau2019ab}, and in Section~\ref{sec:ferminetstar} we describe Fermi Net*, which is functionally identical but better performing (faster) than the original.

At a high level, the Fermi Net consists of
\begin{enumerate}[I]
    \item learnable single electron features (single streams),
    \item learnable electron-electron interaction features (pairwise streams),
    \item permutation equivariant operations (EQV),
    \item and multi-electron orbitals.
\end{enumerate}

To a lesser extent the implementation and optimization details are necessary to the performance and usage. As such they are characteristic of the Fermi Net implementation:

\begin{enumerate}[I]
    \setcounter{enumi}{5}
    \item \gls{kfac} optimization;
    \item and stable log-domain computation of the amplitudes, first order derivatives and second order derivatives.
\end{enumerate}

Next, in this background, we give an overview of the Fermi Net model, describing I-IV. Then we detail \gls{kfac}, describing VI. Details on VII, including the LogSumExp trick and derivations of the derivatives of the determinant, can be found in the appendix of Reference~\cite{pfau2019ab} or understood from other references \cite{bischof2008advances}.

\subsubsection*{Ansatz}

There are two sets of streams in the network referred to as the single and pairwise streams. These streams contain the data corresponding to single, $\mb{h}^{l\alpha}_i$, and pairwise, $\mb{h}_{ij}^{l\alpha\beta}$, electron input features, respectively. These variables are indexed by $l$, the layer of the network, and $i$ and $j$, the electron indexes. $\alpha$ is the spin of electron $i$ and $\beta$ is the spin of electron $j$.

The inputs to the network indexed by $l=0$ and computed from the system $X$ are
\begin{align}
    \mb{h}^{0 \alpha}_i = (&\mb{r}_i - \mb{R}_0, \norm{\mb{r}_i - \mb{R}_0}, \mb{r}_i - \mb{R}_1, \norm{\mb{r}_i - \mb{R}_1}, \nonumber \\ &..., \mb{r}_i - \mb{R}_n, \norm{\mb{r}_i - \mb{R}_n}), \label{eq:fn_singleinputs} \\
    \mb{h}^{0\alpha\beta}_{ij} &= (\mb{r}_i - \mb{r}_j, \norm{\mb{r}_i - \mb{r}_j}). \label{eq:fn_pairwiseinputs}
\end{align}
$\mb{r}_i$ and $\mb{R}_j$ are the electron and atom position vectors, as shown in Figure~\ref{fig:system}, and $\norm{\cdot}$ is the Euclidean norm.

The data from the streams at each layer are transformed by a permutation equivariant function. The permutation equivariant function maintains the anti-symmetry (required by fermionic systems) of the Ansatz and generates multi-electron orbitals with demonstrably good modelling capacity. This function outputs one streams of data $\mb{f}^{l\alpha}_i$:
{\footnotesize 
\begin{equation}
    \mb{f}^{l\alpha}_i = \Bigg( \mb{h}^{l\alpha}, 
    \frac{1}{n_\uparrow}\sum_{\beta \neq \downarrow} \mb{h}^{l\alpha\beta}_{ij},
    \frac{1}{n_\downarrow}\sum_{\beta \neq \uparrow} \mb{h}^{l\alpha\beta}_{ij}, \frac{1}{n_\uparrow}\sum_{\alpha \neq \downarrow} \mb{h}^{l\alpha}_i, \frac{1}{n_\downarrow}\sum_{\alpha \neq \uparrow} \mb{h}^{l\alpha}_i \Bigg)
    \label{eq:fn_pfauequivariant}
\end{equation}}
The data at each layer $l$ are computed
\begin{align}
    \mb{h}^{l\alpha}_i &= \tanh \big( \mb{W}_l\mb{f}^{l\alpha}_i + \mb{b}^l \big) + \mb{h}^{(l-1)\alpha}_i \nonumber \\ \mb{h}^{l\alpha\beta}_{ij} &= \tanh \big( \mb{V}_l \mb{h}^{l\alpha\beta}_{ij} + \mb{c}_l \big) + \mb{h}^{(l-1)\alpha\beta}_{ij} \label{eq:ferminet_layerb} 
\end{align}
where in a layer $l$, $\mb{W}_l$, and $\mb{V}_l$ are weights, $\mb{b}_l$ and $\mb{c}_l$ are the biases. Residual connections are added to all layers where $\mathrm{dim}(\mb{h}^l) = \dim(\mb{h}^{(l-1)})$. 

There are $n_l$ of these parameterized layers. The outputs $\mb{f}^{L\alpha}_i$ are split into spin dependent data blocks. There is a linear transformation to map $\mb{h}^{L\alpha}_i$ to a scalars which are coefficients of the exponentials in the orbitals. Multiple determinants are generated in this way, indexed by $k$, which contribute to the modelling capacity of the network and the elements of the determinants are the product of the coefficient computed by the Fermi Net and the envelopes
\begin{align}
    \phi^{\alpha k}_{ij} = &(\mb{w}^{\alpha k}_{li} \mb{h}_j^{l\alpha} + d^{\alpha k}_{li}) \times 
    \\ & \sum_m \pi^{\alpha k}_{im} \exp( - | \mb{\Sigma}^{\alpha k}_{im} (\mb{r}_j^{\alpha} - \mb{R}_m)| )
    \label{eq:envelopes_pfau}
\end{align}
$\bm{\Sigma}_{im}^{\alpha k}$ control the anisotropic (direction dependent) behaviour of the envelope whereas the inverse exponential ensures the wave function decays to zero when the electrons are large distances from the nuclei. 

The determinants are constructed
\begin{equation}
    \det[\mb{\Phi}^{\alpha k}] = \begin{vmatrix}
\phi_{00}^{\alpha k} & \hdots & \phi_{0n}^{\alpha k} \\ 
\vdots &  & \vdots \\ 
\phi_{n0}^{\alpha k} & \hdots & \phi_{nn}^{\alpha k}
\end{vmatrix}.
\end{equation}
and the amplitudes are computed from these
\begin{equation}
    \psi(X) = \sum_k \omega_k \det[\mb{\Phi}^{\uparrow k}]\det[\mb{\Phi}^{\downarrow k}]. \label{eq:fn_detproduct}
\end{equation}
Equation~\eqref{eq:fn_detproduct} is a representation of the full determinant as a product of spin up and down determinants. This representation forces off-block-diagonal elements of the full determinant to zero, when 
\begin{align}
    & (i \in \{1, ..., n_\uparrow\} \land j \in \{n_\uparrow+1, ..., n\}) \nonumber \\ &\lor (i \in \{n_\uparrow+1, ..., n\} \land j \in \{1, ..., n_\uparrow\}) \nonumber \\
    &= (i \leq n_\uparrow \land j > n_\uparrow) \lor (i > n_\uparrow \land j \leq n_\uparrow),
\end{align}
where $i$ and $j$ refer to the orbital index. Finally, the sign is split from the the amplitudes and the network outputs the amplitudes in the log-domain for numerical stability
\begin{equation}
    \log |\psi(X)| = \log \Big|\sum_k \omega_k \det[\mb{\Phi}^{\uparrow k}]\det[\mb{\Phi}^{\downarrow k}] \Big|
\end{equation}
Other work \cite{spencer2020better}, removes the weights $\omega_k$ in Equation~\eqref{eq:fn_detproduct}, as they are functionally redundant, and replaces the anisotropic decay parameters $\bm{\Sigma}_{im}^{\alpha k}$ in Equation~\eqref{eq:envelopes_pfau} with a single parameter, restricting the orbitals to isotropic decay. Surprisingly, the authors note this does not seem to result in a decrease in modelling accuracy but does lead to significant gains in speed. These improvements are further discussed in Section~\ref{sec:results}.

\subsubsection*{Kronecker Factored Approximate Curvature}

The optimization algorithm used here is a variant approximate natural gradient descent method known as \gls{kfac}. Roughly speaking, updates (the approximate natural gradients) $\tilde{\delta}_l$ for layer $l$ are computed as
\begin{align}
   \tilde{\delta}_l &= \tilde{F}_l^{-1} \text{vec}(\Delta_l\mathcal{L}) \nonumber \\
    &= \bar{A}_l^{-1} \Delta_l\mathcal{L} \bar{S}_l^{-1},
    \label{eq:fn_kfacgradients}
\end{align}
where $\tilde{F}_l^{-1}$ is the approximate inverse Fisher block, $\Delta_l\mathcal{L}$ are the gradients corresponding to weights in layer $l$, and $\bar{A}_l$ and $\bar{S}_l$ are the moving covariances of the left and right Fisher factors, respectively, discussed in literature \cite{martens2015optimizing, grosse2016kronecker, ba2016distributed, martens2018kronecker}. The term Fisher block is used to indicate the elements of the  \gls{fim} corresponding to a given layer in the network. 
In Reference~\cite{pfau2019ab} the authors use a reduced version of the full \gls{kfac} algorithm. They do not use adaptive damping or adaptive learning rates which are somewhat characteristic of the original algorithm \cite{pfauprivate} and considered important in the literature \cite{martens2015optimizing}. 

\begin{figure}
    \centering
    \includegraphics[width=0.5\textwidth]{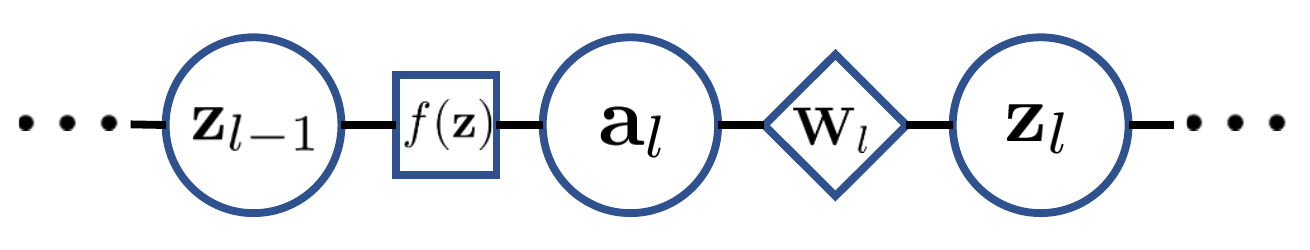}
    \caption{Sketch of the relationship between $\mb{a}$ and $\mb{z}$. For some layer $l$, $\mb{z}_l$ are the pre-activations, $f(\mb{z})$ is an activation function, $\mb{a}_l$ are the activations and $\mb{w}_l$ are the weights. Data variables are in circles, functions in square and network parameters in diamond.}
    \label{fig:nn_sketch}
\end{figure}

The \gls{kfac} variant used in this work is closer to an adaption named Kronecker Factors for Convolution \cite{grosse2016kronecker} because the weights in most of the layers are reused. For example, the single stream weights are used $n_e$ times, where $n_e$ is the number of electrons in the system. There are different approaches to approximating the effect this has on the \gls{fim} and here we use the simpler and more efficient approximation developed in Reference~\cite{ba2016distributed} 
\begin{equation}
    \tilde{F}_l = |T|^2 \EX_X\big[ \EX_i [\mb{a}_{li}]\EX_i [\mb{a}_{li}]^T \big] \otimes \EX_X \big[ \EX_i [\mb{s}_{li}]\EX_i [\mb{s}_{li}]^T \big]
\end{equation}
where $^T$ indicates the transpose, $\mb{a}_{li}$ are the activations of spatial location (the index of data which are operated on by the same parameters) $i$ at layer $l$ and $\mb{s}_{li}$ are the sensitivities of the pre-activations $\mb{z}_{li}$ of layer $l$
\begin{equation}
    \mb{s}_{li} = \frac{d \log |\psi(X)|}{d \mb{z}_{li}}.
\end{equation}
See Figure~\ref{fig:nn_sketch} for a sketch of how these variables are related. $\EX_i$ is the expectation taken over all spatial locations 
\begin{align}
    \EX_i[\mb{a}_{li}] &= \frac{1}{\abs{T}}\sum_i \mb{a}_{li} \\
    \EX_i[\mb{s}_{li}] &= \frac{1}{\abs{T}}\sum_i \mb{s}_{li}
\end{align}
where $T$ is the set of spatial locations and $|T|$ is its cardinality. 

\subsection{Diffusion Monte Carlo}\label{sec:dmc}

Whereas \gls{vmc} relies on the optimization of the parameters of the Ansatz via the derivatives of the expectation value of the energy, \gls{dmc} is a projector method relying on repeated application of the imaginary time operator. Given a trial function, which has some non-zero overlap with the ground sate, repeated application of the Hamiltonian (via the power method) will project the wave function toward the ground state. Green's function Monte Carlo is an example of a projector method \cite{lee1981green}. Diffusion Monte Carlo is a related but distinct projector method that is equivalent to a Green's function method in the limit of small time steps \cite{foulkes2001quantum, maldonado2010quantum, umrigar1993diffusion}. 

Starting with the imaginary time \schro equation
\begin{equation}\label{eq:ml4qc_imaginarytimeschrodinger}
    -\frac{\pd \psi(X, t)}{\pd t} = (\hat{H} - E_T) \psi(X, t)
\end{equation}
that can be interpreted as a diffusion equation and solved in the path integral formalism \cite{kosztin1996introduction} and $E_T$ is some offset. The ground state of the system is a stationary state; there are no time dynamics and the LHS is equal to zero. 

Attempting to simulate the imaginary time \schro equation via the application of the approximated imaginary time operator na{\"i}vely one would encounter the sign problem \cite{gupta2020elucidating, troyer2005computational}, resulting from the antisymmetry constraint of exchange of electrons. The most successful approach to avoiding the sign problem in \gls{dmc} simulations is the fixed node approximation \cite{anderson1976quantum}. In this approximation, the nodes (points in space where the wave function changes from positive to negative) are fixed in place. Practically, this is implemented by not allowing walkers to cross nodes during the sampling process. 

For some state $\psi_n(X) \neq \psi_0(X)$ with eigenvalue $E_n > E_T$ we can see that the derivative of the excited states amplitudes will also be negative: The amplitudes of the states with eigenvalues $E_n > E_T$ decay as a function of time. It is clearer to see when the wave function is expanded as a linear combination eigenfunctions
\begin{align}
    -\frac{\pd \psi(X, t)}{\pd t} = (\hat{H} - E_T) \sum_i \alpha_i \psi_i(X, t) 
\end{align}
and the eigenfunctions of this equation are 
\begin{align}
    \psi(X, t) &= e^{-t(\hat{H} - E_T)}  \sum_{i=0}^\infty \alpha_i  \psi_i(X) \nonumber \\
    &= \sum_{i=0}^\infty \alpha_i  e^{-t(E_i - E_T)} \psi_i(X) 
    \label{eq:ml4qc_schrotimedependentsolns}
\end{align}
in the limit of $t \rightarrow \infty$ the behaviour of the wave function is dependent on $E_T$. There are three cases for the behaviour in the asymptotic region:
\begin{equation}
    \lim_{t \rightarrow \infty} \psi(X, t) = 
    \begin{cases}
    \infty &\text{for  } E_T > E_0 \\
     \alpha_0\psi_0 &\text{for  } E_T = E_0 \\
     0 &\text{for  } E_T < E_0
    \end{cases}
\end{equation}
The trial energy is not known beforehand. It is set adaptively dependent on the behaviour the evolution. In \gls{dmc} $E_T$ is varied dynamically to keep the population of walkers finite.

\begin{figure*}[ht]
    \centering
    \includegraphics[width=\textwidth]{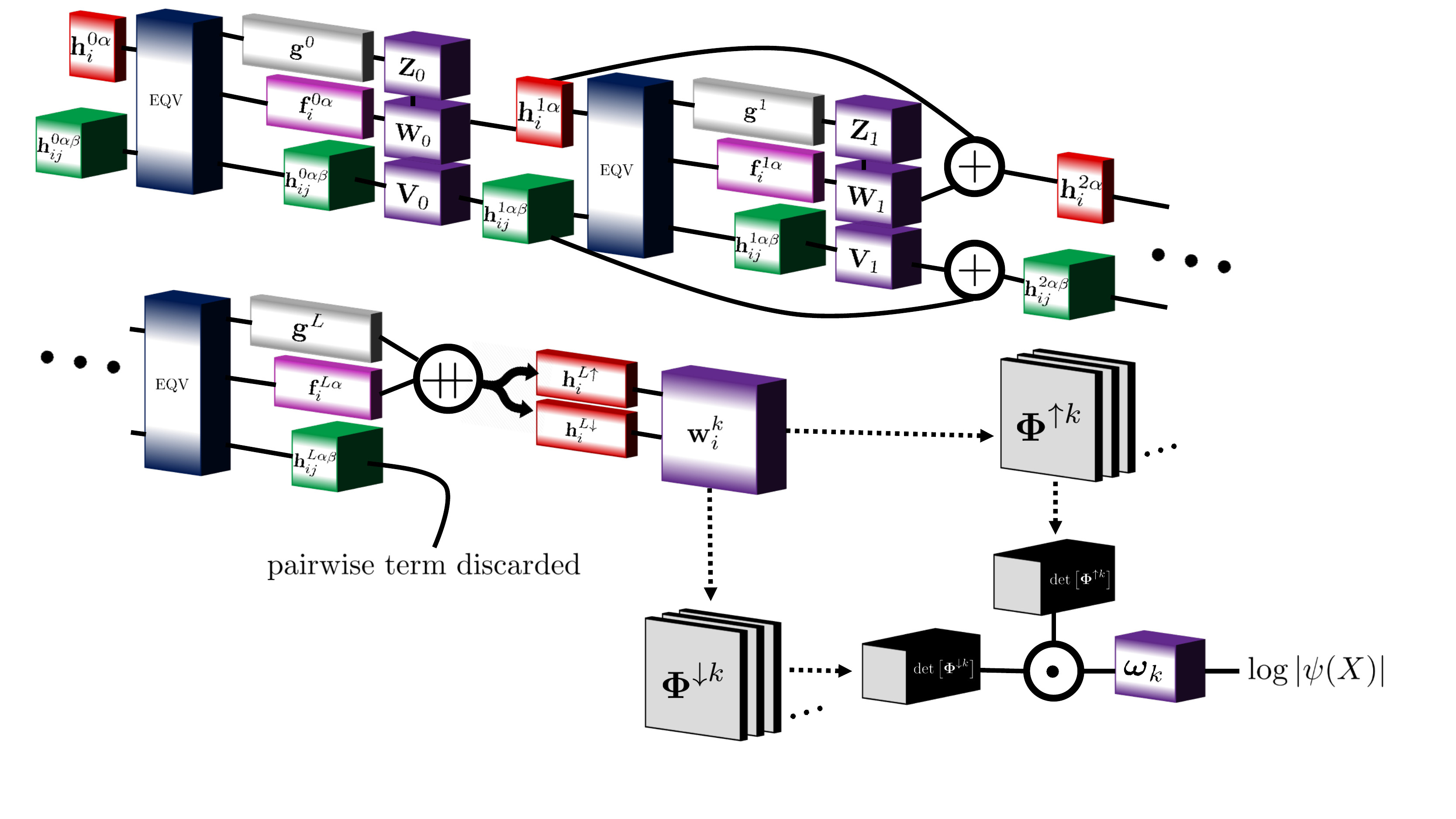}
    \caption{Overview of Fermi Net. The system description $X$ is used to compute the single stream and pairwise stream input feature tensors  $\mb{h}_i^{0\alpha}$ and $\mb{h}_i^{0\alpha\beta}$, respectively. These are passed to the permutation equivariant function (EQV), Figure~\ref{fig:mixer}. Linear layers are applied to the resulting tensors with $\tanh$ activations. Outputs of the Split Stream and Single Stream matrix multiplications are combined before a $\tanh$ activation in the Single Stream layer. These layers are repeated 4 times. After the final permutation equivariant function, the Split Stream and Single Stream outputs are concatenated $(+\!\!\!+)$ and Pairwise Stream data discarded. The concatenated tensor is passed through a final spin dependent linear transformation to spin-up and spin-down determinants. The final layer is a custom computation of the determinants which involves stable first- and second-order derivatives and the LogSumExp trick.}
    \label{fig:ferminet}
\end{figure*}

\section{Methods}\label{sec:methods}
\subsection{Fermi Net*}\label{sec:ferminetstar}

\begin{table}[h]
\begin{tabular}{|l|l|l|}
\hline
Name                   & symbol        & value \\ \hline
Single Stream hidden units & $n_{s}$ & 256 \\
Pairwise Stream hidden units & $n_{p}$ & 32 \\
Split Stream hidden units & $n_{ss}$ & 256 \\
Determinants & $n_{k}$ & 16 \\
Layers & $n_{l}$ & 4 \\
\hline
\end{tabular}
\caption{Model hyperparameters.
\label{tab:model_hyperparameters}}
\end{table}

\begin{figure*}[t]
    \centering
    \includegraphics[width=0.75\textwidth]{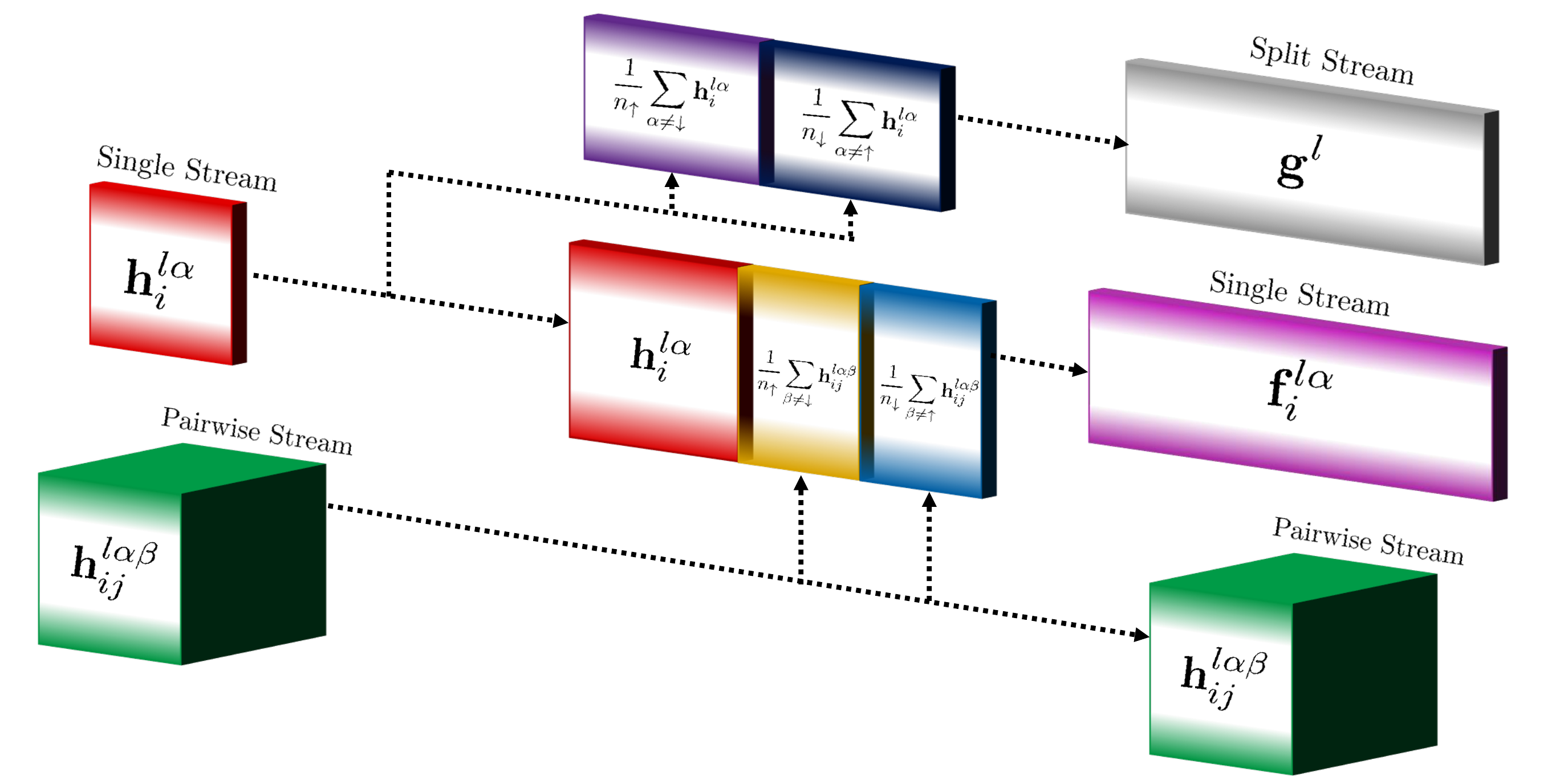}
    \caption{Data from the single and pairwise streams are combined in this operation via Equations~\ref{eq:fn_maxequivarianta} \&~\ref{eq:fn_maxequivariantb}. The pairwise stream data remains unchanged.
    \label{fig:mixer}}
\end{figure*}

We change the network slightly, decreasing the resources (defined as the number of operations) required by the original model. We refer to this implementation of the model (and distinct optimization) as Fermi Net* in order to distinguish from other work, Reference~\cite{pfau2019ab}, which is referred to as Fermi Net. For both implementations of the model we compute the total number of operations of the implementations and compare the walltime.

\begin{figure*}[t]
\centering
    \subfloat[\label{fig:fn_parameterscalinga} Resource comparison of different implementations. The orange line is overlayed by the green line.]{\includegraphics[width=0.45\textwidth]{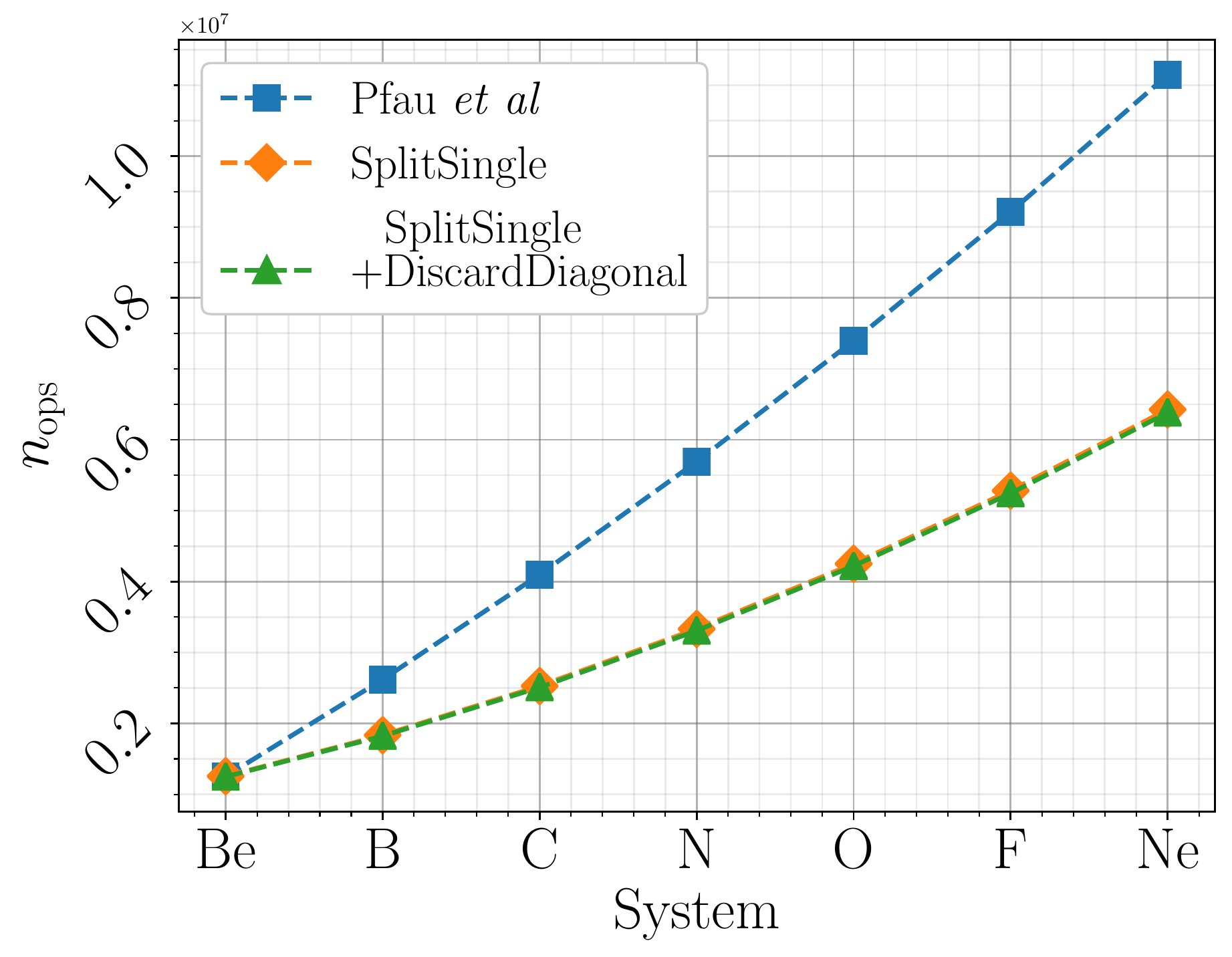}}
    \qquad
    \subfloat[\label{fig:fn_parameterscalingb} Reduction in average walltime resulting from the change of implementation. These values are the average of 1000 sampling steps and 100 energy computations of the model on 1 V100 GPU.]{\includegraphics[width=0.45\textwidth]{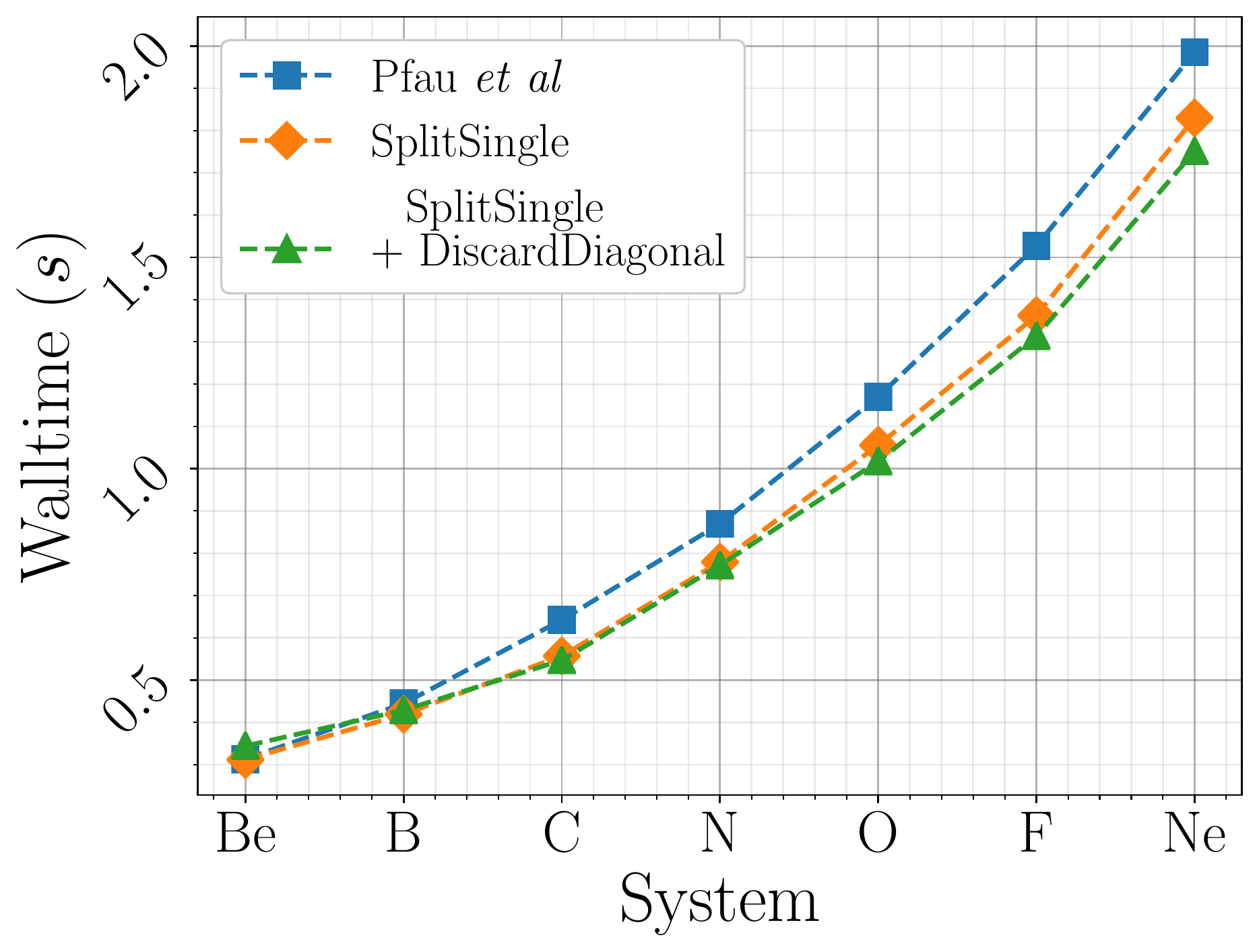}}
    \caption{\label{fig:fn_walltimereduction} The blue line corresponds to the network outlined in Reference~\cite{pfau2019ab}, the orange line to the method of splitting the single streams, Equations~\ref{eq:fn_maxequivarianta} and~\ref{eq:fn_maxequivariantb}, and the green line the resource requirements of splitting the single stream and removing redundant pairwise streams. The resource requirements are measured as the number of required operations, $n_\text{ops}$, Equation~\eqref{eq:fn_nops}. The walltime comparison of these methods is shown in (b). It is important to note that the computational time of the framework is dominated by the determinant calculation. These improvements will become negligible at much larger systems.}
\end{figure*}

The number of operations $n_\mathrm{ops}$ is computed as 
\begin{equation}\label{eq:fn_nops}
    n_\mathrm{ops} = \sum_l n^l_\mathrm{uses} \times d^l_\mathrm{out} \times (d^l_\mathrm{in} + (d^l_\mathrm{in} - 1)) 
\end{equation}
where $l$ is index running over all layers performing matrix multiplications and $n^l_\mathrm{uses}$, $d^l_\mathrm{out}$ and $d^l_\mathrm{in}$ are the number of times those weights are used, the input dimension and the output dimension, respectively. To clarify, $d_\mathrm{out} \times (d_\mathrm{in} + (d_\mathrm{in} - 1))$ is the cost of one matrix-vector product in terms of the number of operations (multiplication or addition). This equation only computes the contributions from the linear layers because these are the only layers where the implementations differ and ignores other operations such as computing the determinant, which is the dominant operation in the complexity.
It is possible to completely remove elements from the pairwise streams tensor with no effect to the modelling capacity or performance of the network. The diagonal elements of the pairwise tensor $\mb{h}^{l\alpha\beta}_{ij}$ where $i=j$ are redundant. The inputs computed from Equation~\eqref{eq:fn_pairwiseinputs} are zero, have zero contribution to computation of the energy (and therefore zero contribution to the computation of the gradients). Though this only results in a small reduction in resource, Figure~\ref{fig:fn_parameterscalingb}, we find a per iteration walltime reduction of $\sim5-10\%$. The effect is more noticeable at larger system sizes. 
The permutation equivariant function is changed to decrease the number of operations, Figure~\ref{fig:mixer}. This new function outputs two streams of data, $\mb{f}^{l\alpha}_i$ and $\mb{g}^l$:
\begin{align}
    \mb{f}^{l\alpha}_i &= \Bigg( \mb{h}^{l\alpha}, 
    \frac{1}{n_\uparrow}\sum_{\beta \neq \downarrow} \mb{h}^{l\alpha\beta}_{ij},
    \frac{1}{n_\downarrow}\sum_{\beta \neq \uparrow} \mb{h}^{l\alpha\beta}_{ij}
    \Bigg) 
    \label{eq:fn_maxequivarianta} \\
    \mb{g}^l &= \Bigg(\frac{1}{n_\uparrow}\sum_{\alpha \neq \downarrow} \mb{h}^{l\alpha}_i, \frac{1}{n_\downarrow}\sum_{\alpha \neq \uparrow} \mb{h}^{l\alpha}_i
    \Bigg).
    \label{eq:fn_maxequivariantb}
\end{align}
This implementation requires less resources than the original, as the mean over spin terms in the layer, $\mathbf{g}^l$, are only operated on once, instead of $n_e$ times, in the case that these data are not split and $\mathbf{g}^l$ appears in all single electron streams.

The data at each layer $l$ are computed
\begin{align}
    \mb{h}^{l\alpha}_i = \tanh&\big(\mb{W}_l\mb{f}^{l\alpha}_i + \mb{Z}_l\mb{g}^{l-1} + \mb{b}^l\big) \nonumber \\ & + \mb{h}^{(l-1)\alpha}_i  \\
    \mb{h}^{l\alpha\beta}_{ij} = \tanh&\big( \mb{V}_l \mb{h}^{l\alpha\beta}_{ij} + \mb{c}_l \big) + \mb{h}^{(l-1)\alpha\beta}_{ij} 
\end{align}
where in a layer $l$, $\mb{W}_l$, $\mb{Z}_l$ and $\mb{V}_l$ are weights, $\mb{b}_l$ and $\mb{c}_l$ are the biases. As before, residual connections are added to all layers where $\mathrm{dim}(\mb{h}^l) = \dim(\mb{h}^{(l-1)})$. The outputs $\mb{f}^{L\alpha}_i$ and $\mb{g}^L$ are concatenated, $\mb{h}^{L\alpha}_i = (\mb{f}^{L\alpha}_i, \mb{g}^L)$, and split into two spin dependent data blocks.

This is an alternate but equivalent representation of the permutation equivariant function outlined in Reference~\cite{pfau2019ab}. This representation reduces the number of operations required to perform a forward pass in the network, Figures~\ref{fig:fn_parameterscalinga} and~\ref{fig:fn_parameterscalingb}.

Using this implementation results in a worse approximation to the \gls{fim}, discussed further in the next section, see Figure~\ref{fig:fisher_layers}, though we did not observe meaningful differences in performance between the two implementations. The reduction in computational effort (as measured by the per iteration walltime as a proxy) was not as large as the drop in resource requirements, though was significant enough ($\sim5-10\%$) to warrant the additional complication.

A complete set of the hyperparameters of the Fermi Net* Ansatz used in these experiments are given in Table~\ref{tab:model_hyperparameters}.

\subsection{Kronecker Factored Approximate Curvature}

We found divergent behaviour of the damping in the asymptotic region of the training, most likely due to the relatively large noise on the loss and small precision requirements of this particular optimization, though more rigorous investigation will likely yield interesting and useful insight.

As such, we use a reduced version of \gls{kfac}. We decay the learning rate, norm constraint and damping
\begin{equation}
    x = \frac{x_0}{1 + t\times10^{-4}}
\end{equation}
where $x$ stands in for the damping, learning rate and norm constraint and $x_0$ for the initial values. $t$ is the iteration of the optimization. The norm constraint is particularly important at the start of training when the quadratic approximation to the local optimization space is large, the natural gradients are large and lead to unstable optimization. In this region the norm constraint plays a role and effectively clips the gradients. Later in the optimization the approximate natural gradients are small and not constrained. The damping / learning rate interplay is an essential ingredient to a functioning \gls{kfac} implementation. The algorithm has nice convergence properties, for this problem especially, but high performance is only possible with careful tuning of the damping and learning rate.

\begin{figure*}
     \subfloat[\label{fig:fisher_layers} Fisher factor $A$ of only the layers containing $\mb{Z}_l$, $\mb{W}_{l}$ and $\mb{V}_{l}$ parameters. $\mb{W}^\prime_{l}$ represents the variables for computing the Fisher factor from a layer where the parameters $\mb{W}_l$ and $\mb{Z}_{l}$ are not split into two layers.]{\includegraphics[width=0.35\textwidth]{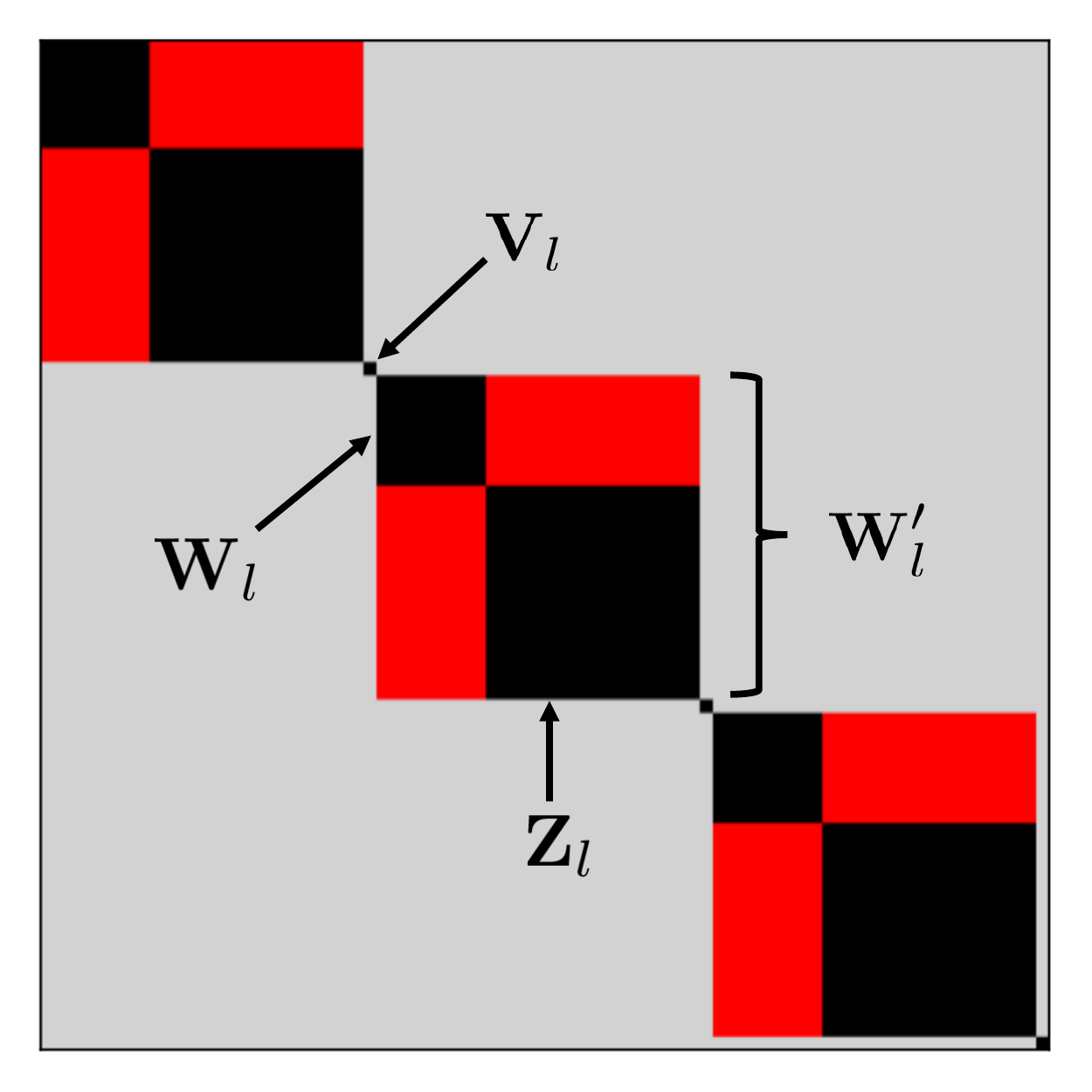}}
     \qquad
     \subfloat[\label{fig:fisher_envs} Fisher factors $A$ of the layers corresponding to the $\bm{\pi}_{im}^{\alpha k}$, $\bm{\Sigma}_{im}^{\alpha k}$ and $\bm{\omega}_k$ parameters.]{
     \includegraphics[width=0.35\textwidth]{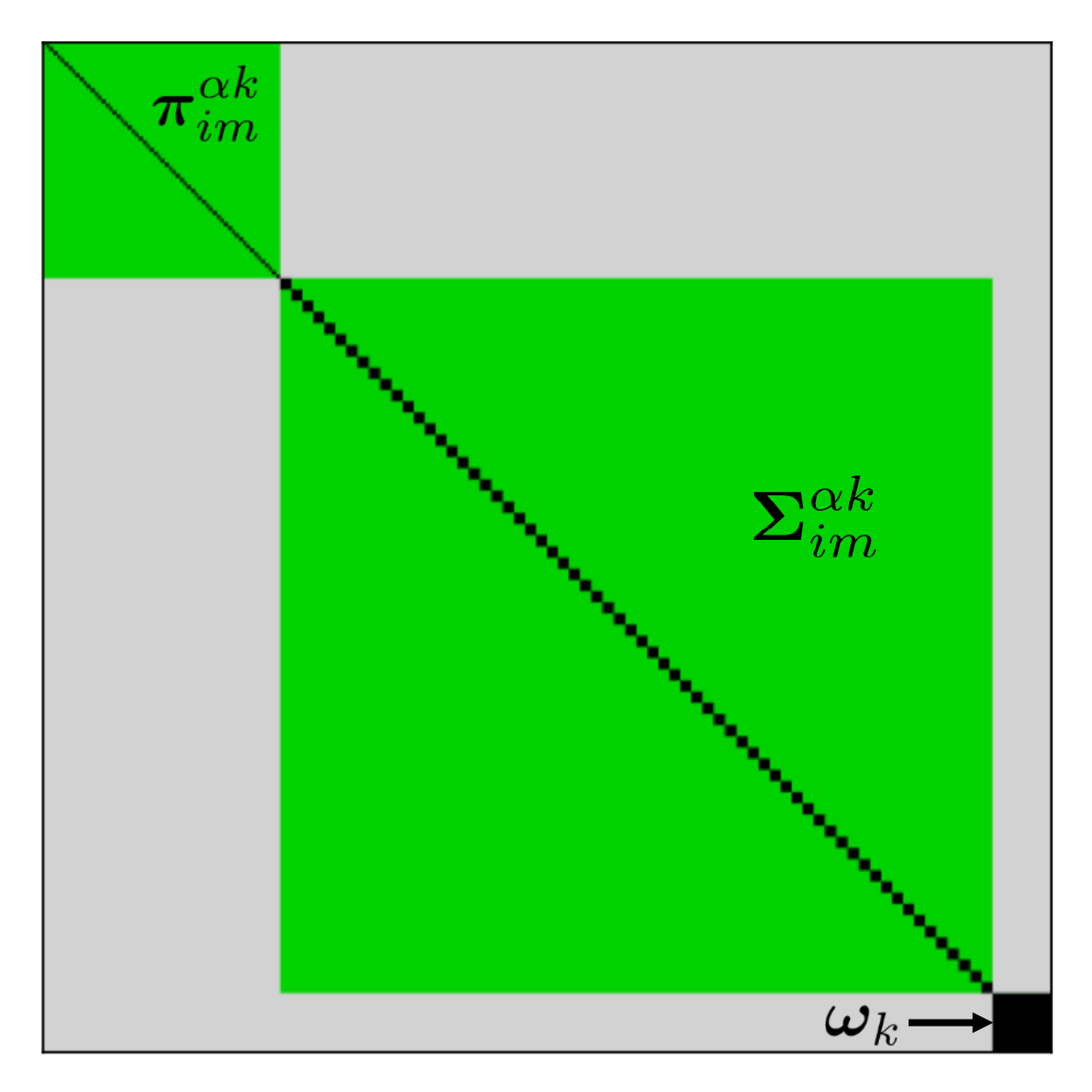}}
    \caption{\label{fig:fisher_comparison} Two cartoons of the differences between the approximations to the `left Fisher factor' $A$, Equation~\eqref{eq:fn_kfacgradients}, in this work and in Reference~\cite{pfau2019ab}. Regions in gray are not computed in either case. Black are computed in both cases. Red are regions lost in this work and green are regions gained. The variable labels (e.g. $\bm{\pi}^{\alpha k}_{im}$) are the parameters corresponding to the Fisher factor.}
\end{figure*}

We bundle parameters in the envelope layers by using sparse matrix multiplications. We compute the entire Fisher block for all parameters of the same symbol (i.e. the $\bm{\pi}$ and $\bm{\Sigma}$ parameters). This is illustrated in Figure~\ref{fig:fisher_envs}. Although we found this implementation was faster in practice, it performs more operations than maintaining a separation between these layers. It does not scale favorably for larger systems, though relative to the computational cost of the network may be negligible.

In Figures~\ref{fig:fisher_layers} and~\ref{fig:fisher_envs} we show the changes on the left Fisher factor resulting from using an implementation which splits data in the permutation equivariant function, versus an implementation which does not. For atomic systems, the right Fisher factor, $\bar{S}$, is the same in this new formulation. Therefore the changes in $\bar{A}$ are representative of the changes to the entire \gls{fim}.

The quality of approximation to the \gls{fim} is dependent on the data distribution and the model. In this problem and for a small model, we found that this approximation provided a better representation of the exact \gls{fim} than other methods \cite{grosse2016kronecker} when compared on a small model over an initial optimization run of $1 \times 10^3$ iterations. It is not possible to perform the comparison to the exact \gls{fim} as the memory requirements scale as $\mathcal{O}(n_\text{w}^2)$, where $n_\text{w}$ is the number of parameters. 

\subsection{Variational Monte Carlo with Kronecker Factored Approximate Curvature}

The goal is to create a good enough approximation of the wave function and therefore the nodal structure of the wave function using \gls{vmc} to be followed up with \gls{dmc} given the Fermi Net* Ansatz. The \gls{vmc} requires $c_l$ forward passes (sampling), 1 backward pass (gradient) and 1 energy computations. To sample the wave function we use Metropolis Hastings. The step size is adaptively changed during training to move the acceptance toward the target sampling acceptance ratio of $0.5$. $c_l$ is the correlation length and here was set to $10$. The model is pretrained for $1\times10^3$ iterations using the methods outlined in Appendix~\ref{app:pretraining}.

The gradients of the parameters are computed from Equation~\eqref{eq:vmc_gradients} and the centered energies $(E_L(X) - \EX_X[E_L(X)])$ in Equation~\eqref{eq:vmc_gradients} have been clipped 5x from the median value. Walkers $X$ are sampled via Metropolis Hastings. A walker is a point in the configuration space that moves around the space by taking random steps. These moves are accepted or rejected dependent on the ratio of in probability of the starting and end points. A description of the algorithm is given in Algorithm~\ref{alg:metropolishastings}.

\begin{algorithm}[H]
\caption{Metropolis-Hastings algorithm used here for sampling. $c_l$ is the correlation length. $\mb{r}$ and $xi$ are $M \times n_e \times 3$ dimensional tensors. Steps update electron positions simultaneously and acceptance of steps are performed in parallel across all walkers. Line 7 updates walkers where the condition is true. $\sigma$ is the step size that is adaptively determined to maintain an acceptance ratio of 0.5. \label{alg:metropolishastings}}
\begin{algorithmic}[1]
\For{$c_l$}
    \State $\xi \sim N(0, \sigma)$
    \State $\mb{r}^\prime \gets \mb{r} + \xi$
    \State $\alpha \sim U[0, 1]$
    \State $P_\mathrm{move} \gets \frac{p(\mb{r}^\prime)}{p(\mb{r})}$
    \If{$P_{\mathrm{move}} > \alpha$}
        \State $\mb{r} \gets \mb{r}^\prime$
    \EndIf
\EndFor
\end{algorithmic}
\end{algorithm}

The approximate natural gradients for layer $l$, which are the updates to the Ansatz parameters, are computed via Equation~\eqref{eq:fn_kfacgradients}. The approximation to the Fisher factors, $\bar{A}$ and $\bar{S}$, are `warmed-up' by taking 100 steps of stochastic gradient descent with small learning rate ($\nu = 1\times10^{-5}$) whilst accumulating statistics. This ensures a smoothed initial approximation to the \gls{fim} and is consistent with other work in the literature \cite{martens2015optimizing}.

For all systems, other than Beryllium, $1 \times 10^5$ iterations of \gls{vmc} were run. The full \gls{vmc} and \gls{kfac} algorithm used in this work is outlined in Algorithm~\ref{alg:vmckfac} and a full set of hyperparameters and initialization values of variables are given in Table~\ref{tab:kfacvariables}.

\subsection{Diffusion Monte Carlo}

\begin{figure}
    \centering
    \includegraphics[width=0.5\textwidth]{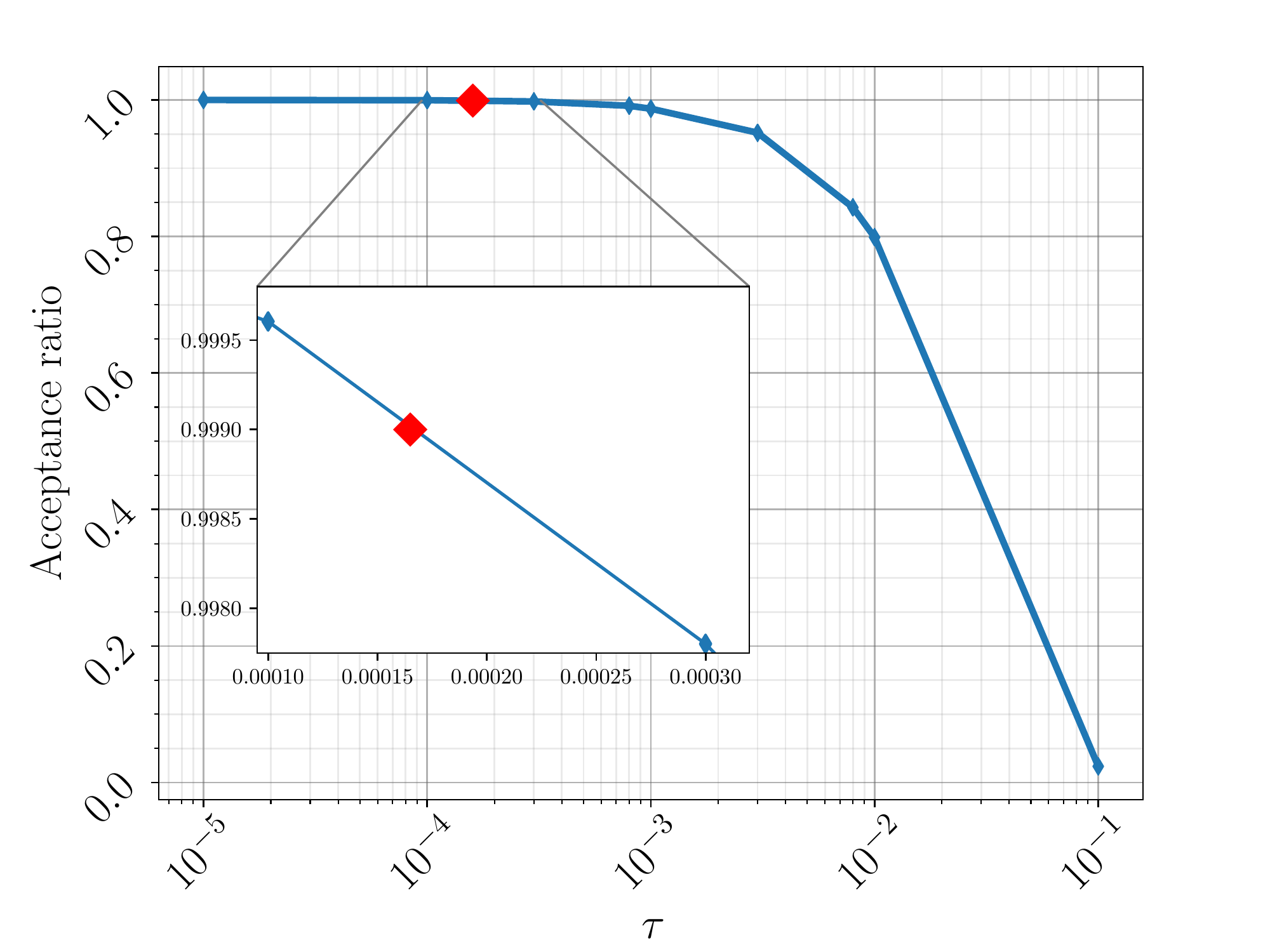}
    \caption{The Acceptance ratio of Nitrogen as a function of the time-step $\tau$. The red diamond marker indicates the estimated value of $\tau$ at an acceptance ratio of 0.999 ($99.9\%$).}
    \label{fig:find_tau}
\end{figure}

The full algorithm used in this work is outlined in Algorithm~\ref{alg:dmc} (found in Appendix~\ref{app:dmc}. This implementation is a simplified version of the algorithm described in Reference~\cite{umrigar1993diffusion}. We make simplifying changes by removing transformations to spherical coordinate systems and set $\tau_\text{eff} = \tau$. Also, we move all walkers and electrons simultaneously with no change to how the algorithm behaves. We set $\tau$ by testing a non-uniform range of values for $\tau$, computing the average acceptance over 100 iterations, and solving the equation of a line between two points to find the closest value that gives an acceptance ratio (Acceptance) of $0.999$, shown in Figure~\ref{fig:find_tau}. We found this worked well in practice and all $\tau$ acceptances were $\sim99.9\%$. A complete list of the hyperparameters needed given in Table~\ref{tab:dmc_variables}.

The \gls{dmc} implementation is parallelized over all electrons and walkers. The model weights and walker configurations are converted to 64-bit floats for the \gls{dmc}. \gls{dmc} was run for a variable number of iterations, where the minimum was $5 \times 10^4$, and this is discussed further in Section~\ref{sec:conclusions}. 

\subsection{Code and Hardware}

\begin{table*}
\centering
\begin{tabular}{c|cc|ccccc}
\hline \hline
Atom & Fermi Net* + DMC & Fermi Net* & Fermi Net \cite{pfau2019ab}    & VMC \cite{seth2011quantum} & DMC \cite{seth2011quantum} & HF & Exact \cite{chakravorty1993ground}     \\ 
& (This work) & (This work)&&&&& \\ 
\hline
Be   & -14.66734(2) & -14.66726(1) & -14.66733(3) & -14.66719(1)           & -14.667306(7) & -14.35188 & -14.66736  \\
B    & -24.65384(9) & -24.65290(2) & -24.65370(3)          & -24.65337(4)  & -24.65379(3)  & -24.14899 &  -24.65391 \\
C$^+$& -37.43086(6) & -37.43061(1) & -37.4307(1)$^\dagger$& -37.43034(6)  & -37.43073(4) & -36.87037 & -37.43088   \\
C    & -37.84472(7) & -37.84452(1) & -37.84471(5) & -37.84377(7) & -37.84446(6)  & -37.08959 & -37.8450   \\
N    & -54.5891(5)  & -54.58755(6) & -54.58882(6) & -54.5873(1)  & -54.58867(8)  & -53.5545  & -54.5892   \\
O    & -75.0667(3)  & -75.0599(2)  & -75.06655(7) & -75.0632(2)  & -75.0654(1)   & -73.6618  & -75.0673   \\
F    & -99.7332(5)  & -99.7277(1)  & -99.7329(1)  & -99.7287(2)  &  -99.7318(1)  & -97.9865  & -99.7339   \\
Ne   & -128.9370(3) & -128.9351(1) & -128.9366(1) & -128.9347(2) & -128.9366(1)  & -126.6045 & -128.9376  \\ \hline \hline
\end{tabular}
\caption{Comparison of \gls{vmc} ad \gls{dmc} results with existing works. Fermi Net* energies are computed from $1 \times 10^4$ batches of size 8096 given a model after $1\times10^5$ of training. The number in the brackets indicates the error on the calculation at the same precision as the value reported. For example, -14.66734(2) indicates a value of -14.66734$\pm$0.00002. All errors reported in this work are the standard error on the mean. \\
$^\dagger$ Fermi Net carbon cation result not directly reported. Computed from ionisation energy and carbon energy result and the errors propagated via addition.}
\label{tab:atomic_results}
\end{table*}

Each \gls{vmc} experiment used 2 V100 GPUs, the \gls{dmc} runs were distributed over a variable number of CPUs due to constraints on the availability of GPU time. 

The code was written in PyTorch 1.5 using CUDA 10.1. The implementation was parallelized such that each GPU held 2 models, or each CPU 1 model. Initial attempts at producing this model were written in Tensorflow 2 with CUDA 10.1 but we experienced significant and sometimes not (easily) diagnosable issues. A simplified version of this Ansatz will be released to DeepQMC \cite{deepqmc}.

Computations of the Hartree Fock orbitals were performed using PySCF \cite{PYSCF} and the model was distributed using Ray \cite{ray}.

\section{Results and Discussion}\label{sec:results}

\subsection*{Diffusion Monte Carlo}

\begin{figure*}
    \centering
    \includegraphics[width=0.9\textwidth]{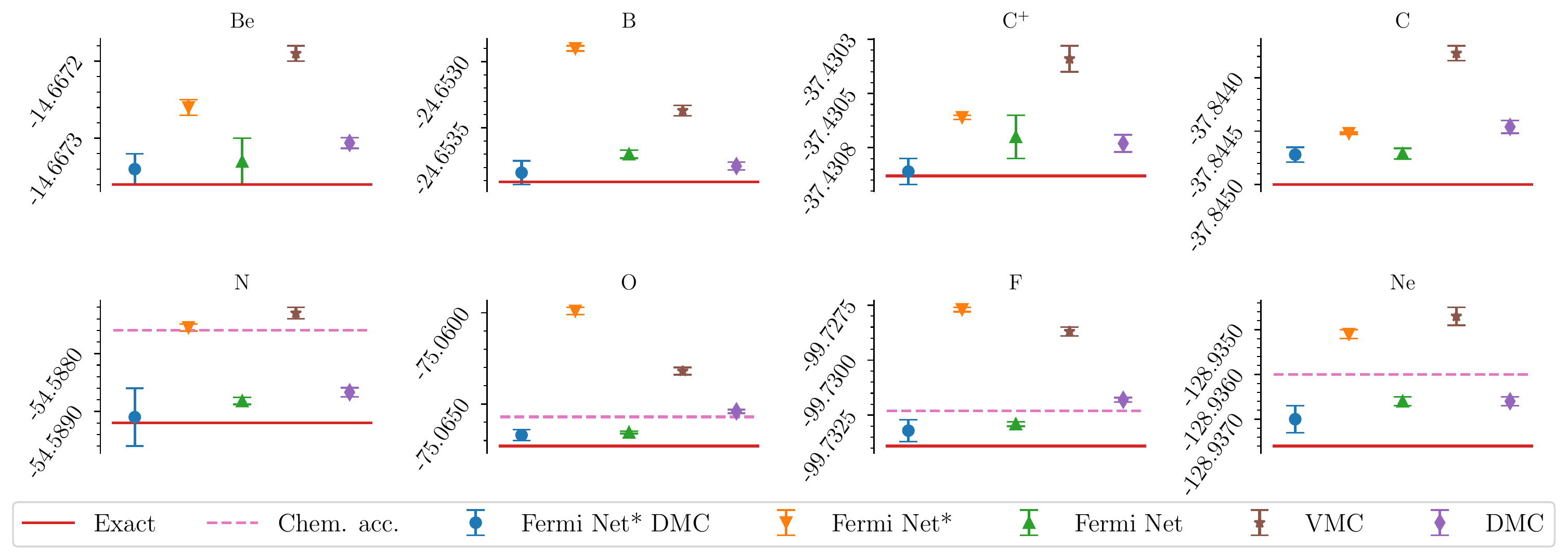}
    \caption{Graphical representation of data from Table~\ref{tab:atomic_results}, see caption for details. A chemical accuracy line (Chem. Acc. - pink dashed line) is plotted where it falls within the range of the plotted data for a system. \label{fig:atomic_results_graph}}%
\end{figure*}

We demonstrate functionality of the introduced computational techniques on the atomic systems from second period (Be-Ne), and the cation C$^+$. For each system, first we optimize parameters of the Fermi Net*, in order to generate a good trial wave function, that is subsequently improved through \gls{dmc} method. As it was demonstrated in \cite{pfau2019ab} Fermi Net is capable of outperforming other \gls{vmc} methods. Our \gls{vmc} results do not exceed the existing state-of-the-art, the \gls{dmc} either exceed or match other best results (see Table~\ref{tab:atomic_results}).


Table~\ref{tab:atomic_results} summarizes the results. All references to Fermi Net* Ansatz indicate the methods outlined in this work. The first column of this Table contains the final energies and errors of the wave function after both \gls{vmc} and \gls{dmc} with the Fermi Net* Ansatz and the second column are the corresponding energies only after the \gls{vmc}. The third through seventh columns (Fermi Net, VMC, DMC, HF, and Exact) contain benchmark results from the literature indicated. The HF column are the Hartree-Fock energies obtained using the STO-3g basis, the basis used in this work for the pretraining orbitals. 

\begin{figure*}
    \centering
    \includegraphics[width=0.9\textwidth]{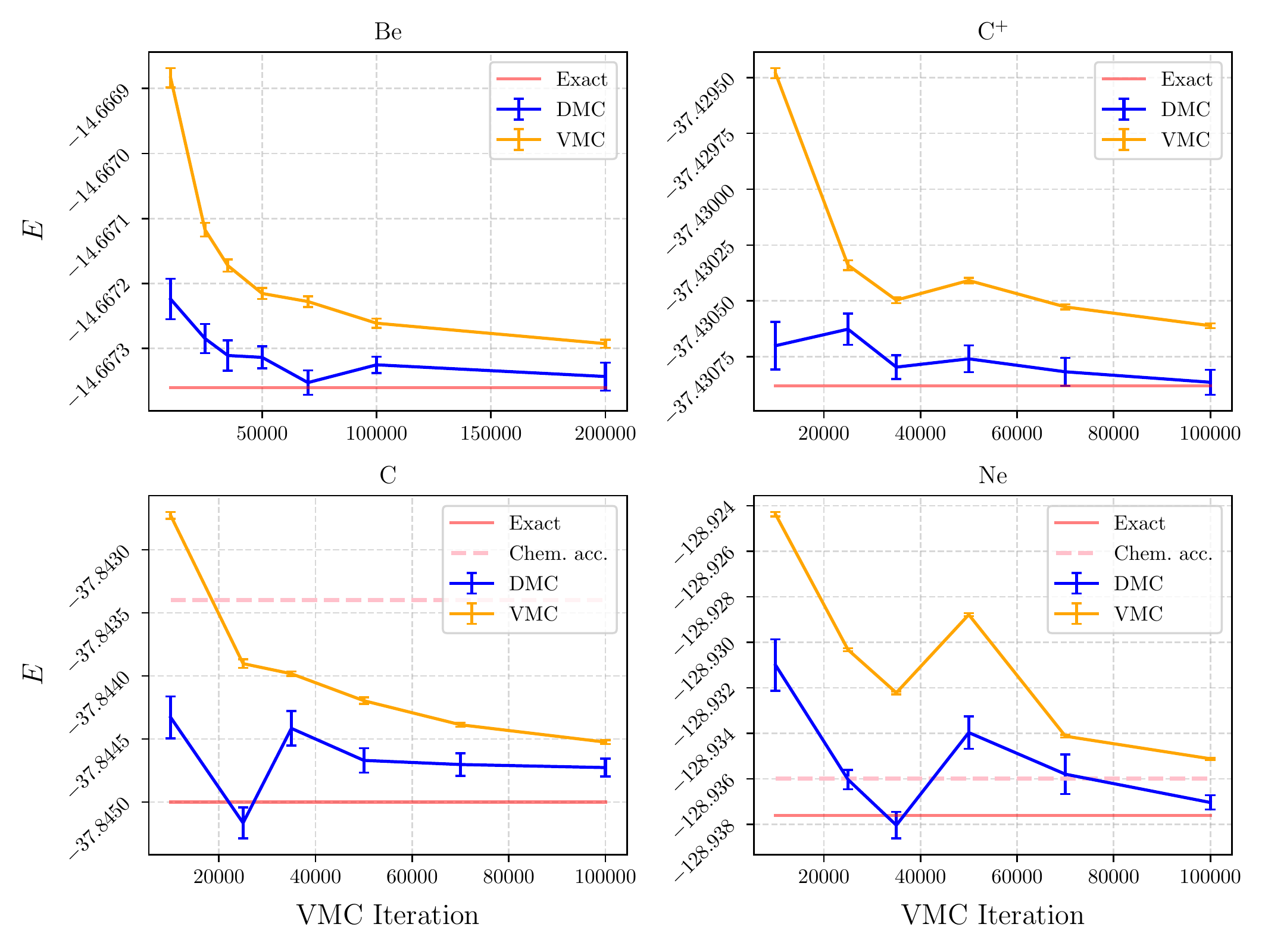}
    \caption{\gls{dmc} applied to Ansatz at different stages of \gls{vmc} optimisation. The red line is the exact energy of the ground state, given in Table~\ref{tab:atomic_results}. The orange line is the energy of the wave function at a particular \gls{vmc} iteration. Note that Be extends to $2\times10^5$. It is demonstrated here to show that there is still capacity for the wave function to improve, but we were restricted by computational time for the other systems. The blue line is the resulting energy from the application of \gls{dmc} to that iteration of the wave function. The \gls{dmc} converged generally after around $5 \times 10^3$ iterations, and was run for between $5\times10^4 - 2\times10^5$ iterations after convergence. The error bars are the standard error $\sigma_\text{SEM} = \sigma / \sqrt{m}$ where $\sigma$ is the standard deviation of the energies of the batches evaluated and $m$ is the number of batches. A chemical accuracy line (pink dashed) is added to the plots where it is within the range of the plotted data.}
    \label{fig:vmc_dmc}
\end{figure*}

The Fermi Net* \gls{vmc} energies in column 2 are not directly comparable to the Fermi Net energies. We focused computational resources on the \gls{dmc} and restricted all but one system (Be) to $1\times10^5$ iterations, whereas the original Fermi Net work used double this number. We extended one Be to $2\times10^5$ iterations and found improvements to the energy, Figure~\ref{fig:vmc_dmc}, computed as -14.66730(1). This is at the top end of the error in the original Fermi Net result, indicating that the performance may be matched with this implementation, if slightly worse. Highlighting this result clearly does not guarantee equivalent performance on larger systems and more extensive trials on larger systems are required. The other Fermi Net* results are consistently poorer than Fermi Net, but better than the referenced benchmark \gls{vmc} results. 

All energies in column 1 are to within less than 0.25\% of the respective correlation energies, at best (Be) within 0.03\%. Be, B, C$^+$, and N are all accurate to the exact ground state energy within error bars. Systems Be, B, C$^+$ and Ne are the most accurate energies. Systems Be, C$^+$, C and Ne were run with around $2\times10^5$ iterations, which are explored in more depth in Figure~\ref{fig:vmc_dmc}, and B, C, N, O with around $5\times10^4$ iterations of \gls{dmc}. 

The error bars are the standard error $\sigma_\text{SEM} = \sigma / \sqrt{m}$ where $\sigma$ is the standard deviation of the energies of the batches and $m$ is the number of batches evaluated. They are noticeably larger in column 1 of Table~\ref{tab:atomic_results} for the systems mentioned where less iterations of \gls{dmc} are run. \gls{dmc} consistently improved the wave function, that is the evaluated energy of the wave function was lower, but the error bars are significantly larger in some cases, due to the autocorrelation of the data trace. 

Figure~\ref{fig:vmc_dmc} shows \gls{dmc} applied to different trial wave functions. Each \gls{dmc} point is evolved from the corresponding \gls{vmc} iteration. The \gls{vmc} lines in general show a clear trend for improving the energy of the wave function as the iterations increase, and this is mirrored in the energy computed from the \gls{dmc}. There are several apparent anomalies in the data. In the Neon \gls{vmc} line there is a clear decrease in the quality of the model (increase in the energy) during the training. We believe this is a result of divergence in training and may indicate that our implementation of \gls{kfac} becomes more unstable as the molecules become larger. We cannot extrapolate significant conclusions without more analysis of the optimization. In both the Neon and Carbon \gls{dmc} lines there is a large kink in the energy achieved by \gls{dmc}. Further investigation did not reveal divergent behaviour in the \gls{dmc}, that is large changes in energy after convergence, and repeated runs showed more stable behaviour in line with what would be expected. It is possible on these particular runs the walkers became trapped near the nodes and accumulated anomalous low energy statistics. Again, further analysis and development of this \gls{dmc} is required to understand if this is a feature of the precise Ansatz or issues with the \gls{dmc} implementation. 

\subsection{GPU, CPU and Computational Time}

In this work we have used a combination of GPU and CPU compute resources. Typically, research groups may not have access to state-of-the-art compute architectures and may need to exploit alternate compute resources. The GPU methods were significantly faster than the CPU experiments, especially in the cases of larger systems, for example GPU vs. CPU per iteration times for the \gls{dmc} algorithm on the system Neon were ~4s and ~24s, respectively, we found that in these small systems a CPU implementation distributed over 3 nodes was enough for reasonable experiment time ($<$ 1 week). However, the scaling makes the CPU implementation impractical for larger systems. CPU implementations are not standard for the machine learning community, but are for the quantum chemistry community. We highlight here that these methods can be used on CPU architectures, though may quickly become impractical for larger systems. 

One iteration of \gls{dmc} is less computationally demanding than \gls{vmc}, requiring 1 forward pass, 1 energy computation and 1 backward pass, plus some negligible functions.  However, we port the Fermi Net weights to 64-bit precision for the \gls{dmc} phase, finding significantly better performance. This is increases the walltime of 1 iteration of \gls{dmc} to 1-1.5x 1 iteration of \gls{vmc}.

\section{Conclusions}\label{sec:conclusions}

In this work we have changed the structure of a neural network Ansatz, the Fermi Net, by removing redundant elements (the diagonal elements of the pairwise streams) and splitting the data in the permutation equivariant function such that it is not reused unnecessarily. These changes increase the performance (as measured by the walltime) of the network. Additionally, we have improved approximations to the ground state, found with Variational Monte Carlo and a Fermi Net* Ansatz, with Diffusion Monte Carlo, matching or exceeding state-of-the-art in all systems. 

With respect to the first contribution, although this model is small compared with other state-of-the-art neural networks \cite{brown2020language}, it contains the expensive determinant computation, which is the dominant term in the complexity of the network scaling as $\mathcal{O}(n_e^3k)$, where $n_e$ is the number of electrons and $k$ is the number of determinants. Variational Monte Carlo requires computation of the Laplacian, which uses $n_e$ backward passes of the first order derivatives. In total, the estimated complexity of the network is $\mathcal{O}(n_e^4k)$. Although the changes made here improve the performance, this improvement in speed may be negligible at larger system sizes. 

\subsection{Related Work}

Other more recent work \cite{spencer2020better} additionally found significant efficiency gains in altering the neural network with no noticeable reduction in accuracy: replacing the anisotropic decay parameters with isotropic decay parameters in the envelopes; and removing more redundant parameters in the determinant sum.
Further, other smaller networks employing a more traditional Jastorw/backflow Ansatz \cite{hermann2019deep} and integrated Hartree-Fock orbitals were significantly less computationally demanding at the cost of notably worse performance. There are clearly still gains to be made in the efficiency of these methods, and a better understanding on the relationship between the problem and the size of the network required will be important pieces of understanding for tuning these methods.

\subsection{Future Work}

In this work we alternate between 32-bit and 64-bit computational precision for the Variational Monte Carlo and Diffusion Monte Carlo methods, respectively. It may be possible to exploit this further. One example is mixed precision algorithms and networks \cite{micikevicius2017mixed}. A suitable and easy first step might be mixed precision schedules. Using low precision weights earlier in training and switching to high precision in the asymptotic region of the optimization might encourage even better results, for example improving the approximation to the Fisher Information Matrix in the region where there are small variations in the optimization landscape. In early tests, reducing the algorithm and model to 16-bit precision proved unstable by generating singular Fisher blocks (that could not be inverted). 

There is a lot of room to improve the optimization. \gls{kfac} is a powerful algorithm which requires careful tuning. The methods used in this work are relatively simple, notably the schedule of the learning rate and damping, and not representative of the suggested mode of operation. Stable adaptive techniques will be essential to improving the optimization in this domain. Additionally, the efficiency of \gls{kfac} can be improved further by using intuitions about the \gls{fim}. For example, the \gls{fim} will change less in the asymptotic region of the optimization, schedules which exploit this fact and update the Fisher blocks and their inverses less should be used, as outlined in the literature \cite{martens2015optimizing}. 


Finally, there are existing approximations from \gls{vmc} literature which can be easily integrated into these frameworks, such as pseudopotentials (for scaling to larger systems) \cite{quinn2006pseudopotentials}, and alternate representations of the wave function \cite{bajdich2006pfaffian, casula2003geminal}. Additionally, these new and highly precise techniques can be applied to other systems including solids and other interesting Hamiltonians.



\subsection{Final comment}

We conclude that though these results are comparatively good in isolation, the prospects for improvements in these methods and hardware paint a strong future for the application of neural networks in modelling wave functions in the continuum.

\section*{Acknowledgements}
We thank David Pfau for helpful discussions. We are grateful for support from NASA Ames Research Center and from the AFRL Information Directorate under grant F4HBKC4162G001.  F.W.
was supported by NASA Academic Mission Services,
contract number NNA16BD14C. 

\clearpage

\section*{Appendices}
\appendix

\section{Pretraining}\label{app:pretraining}

The pretraining was performed very similarly to as described in Reference~\cite{pfau2019ab}. We outline the methods here for completeness.

The pretraining loss is 
\begin{align}  
    \mathcal{L}^\text{pre} =& \int \bigg[ \sum_{\alpha\in\{\uparrow, \downarrow\}} \sum_{ijk} \Big(\phi^{k\alpha}_{ij} (X) \nonumber \\ &- \phi^\text{HF}_{i\alpha}(\mb{r}^\beta)\Big)^2  \bigg] p^\text{pre}(X) dX
\end{align}
where
\begin{equation}
    p^\text{pre} = \frac{1}{2} \Bigg( \prod_{\alpha\in\{\uparrow, \downarrow\}} \prod_i \phi_{i\alpha}^\text{HF} (\mb{r}^\beta) + |\psi(X)|^2 \Bigg),
\end{equation}
with $\phi_{i\alpha}^{\text{HF}}$ being Hartee-Fock orbitals obtained with STO-3g basis evaluated with PySCF python package. 

This quantity is approximated by splitting the samples (at random) into two equal length sets and taking one Metropolis Hastings step with $p(X) = \prod_{\alpha\in\{\uparrow, \downarrow\}} \prod_i \phi_{i\alpha}^\text{HF} (\mb{r}^{\alpha i})$ for the first set and $p(X) = |\psi(X)|^2$ for the second set. The sets are joined and the process repeated. This implementation is slightly different to the original work, and seems to improve the pretraining: The Ansatz (after pretraining) has a lower energy. 

As described in Reference~\cite{pfau2019ab}, this distribution samples where the \gls{hf} orbitals are large, and also where the wave function is poorly initialized and is incorrectly large. This method of pretraining is a better solution to the problem of guiding the initial wave function. An Ansatz close to the \gls{hf} orbitals seems to not get stuck in local minima and drastically improves the optimization behaviour. Other solutions  include integrating the \gls{hf} orbitals into the network throughout the whole training, for example in Reference~\cite{hermann2019deep}. The optimization was performed using an Adam optimizer with default hyperparameters, notably learning rate $1 \times 10^{-3}$.

\section{Kronecker Factored Approximate Curvature}\label{app:kfac}

\begin{figure*}
    \centering
    \includegraphics[width=0.7\textwidth]{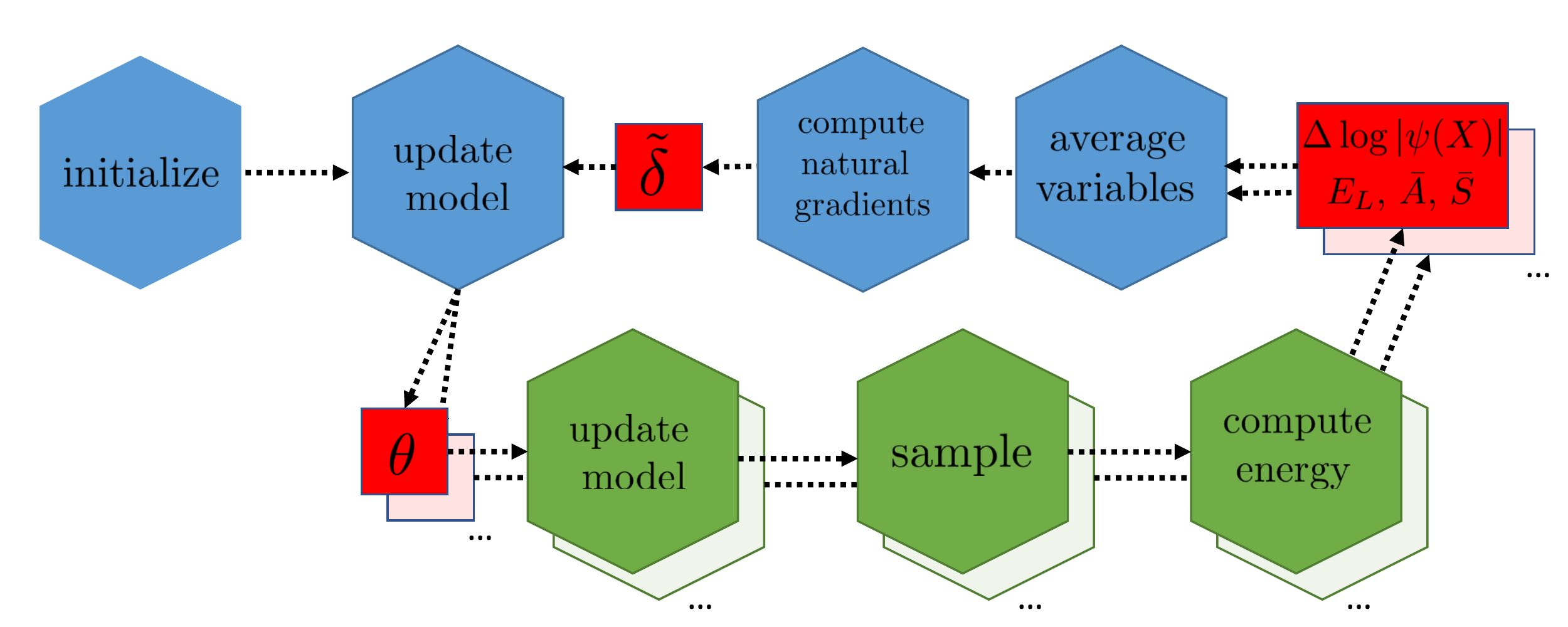}
    \caption{Flow diagram of distributed \gls{kfac} used in this work. The blue hexagons are the head worker, green hexagons workers. There are multiple workers, indicated by the staggered images and ellipses. The red squares are variables computed at one step and used at the next. $\theta$ are the parameters of the model, $\tilde{\delta}$ are the approximate natural gradients, $\Delta \log|\psi(X)|$ are the derivatives of the wave function wrt the electron position vectors and $E_L$ are the local energies of the walkers $X$. These are tensors with $M$ copies, where $M$ is the batch size on a worker. $\bar{A}$ and $\bar{S}$ are the left and right Fisher factors computed during the forward and backward passes. These variables are averaged before being used to compute the approximate natural gradients. Note that the average $\bar{A}$ and $\bar{S}$ are \textit{not} passed back to the workers. This results in a worse approximation to the Fisher, but a difference in performance was not noticed during tests.}
    \label{fig:kfac_distribution}
\end{figure*}

Natural gradient descent \cite{amari1997neural}, is an algorithm for computing the natural gradients of a parameterized function. They are 
\begin{equation}
    \mb{\Delta} = F^{-1} \mathrm{vec}(\Delta\mathcal{L}) \label{eq:natural_gradients}
\end{equation}
where $\mb{\Delta}$ are the natural gradients, $\Delta \mathcal{L}$ are the gradients of the function parameters, $\mathrm{vec}(M)$ is the vectorization of a matrix $M$, and $F$ is the \gls{fim}
\begin{equation}
     F = \EX_X \bigg[\frac{d\log p_\theta(X)}{d \theta}^T \frac{d\log p_\theta(X)}{d \theta}\bigg],
\end{equation}
where $p_\theta(X)$ is some distribution parameterized by $\theta$. \gls{kfac} is an algorithm that approximates the natural gradients of a neural network. There are a series of approximations and methods associated with this algorithm, which are often applied in a case dependent way. The main drawback of Natural Gradient Descent is the cost of inverting the \gls{fim}, which for an $n \times n$ matrix is $\sim O(n^3)$. This is prohibitive for neural networks with even a few thousand parameters. \gls{kfac} reduces this burden whilst still accurately (enough) modelling the Fisher such that useful approximate natural gradients can be found.

Given that the inverse of a Kronecker product is the Kronecker product of the inverses
\begin{equation}
    (A \otimes B)^{-1} = A^{-1} \otimes B^{-1} \label{eq:kronecker_inverse_identity}
\end{equation}
the authors break down the \gls{fim} into blocks, extract the most important blocks, and approximate these blocks in a form that allows the identity of inverting Kronecker products, Equation~\eqref{eq:kronecker_inverse_identity}, to be used. 

Expressing the \gls{fim} as blocks it is natural to use the structure of the neural network to dictate the layout of the blocks: 
\begin{equation}
F = 
\left (
\begin{array}{ccc}

\begin{array}{l} 
\EX[D\theta^{(0)} D\theta^{(0)T}]
\end{array}

& \cdots & 

\begin{array}{l} 
\EX[D\theta^{(0)} D\theta^{(L)T}]
\end{array} \\

\vdots & \ddots & \vdots\\

\begin{array}{l} 
\EX[D\theta^{(0)} D\theta^{(L)T}]
\end{array} &

\cdots & 

\begin{array}{l} 
\EX[D\theta^{(L)} D\theta^{(L)}]
\end{array}

\end{array}
\right )
\end{equation}

where

\begin{equation}
    D\theta^{(i)} = \frac{d\log p_\theta(X)}{d \theta^{(i)}}
\end{equation}
the $^{(i)}$ superscript denotes the parameters in layer $i$ of the network.

Removing all but the diagonal blocks we have
\begin{equation}
\tilde{F} = 
\left (
\begin{array}{ccc}

\begin{array}{l} 
\EX[D\theta^{(0)} D\theta^{(0)T}]
\end{array}

& \cdots & 

0 \\

\vdots & \ddots & \vdots\\

0 &

\cdots & 

\begin{array}{l} 
\EX[D\theta^{(L)} D\theta^{(L)}]
\end{array}

\end{array}
\right )
\end{equation}
Each of these blocks can be reformulated as the Kronecker product of the outer product of the activtations and sensitivities, for example layer $0$,
\begin{align}
    F^{(0)} &= \EX_X[D\theta^{(0)} D\theta^{(0)T}] \nonumber\\
    &= \EX_X[(\mathbf{a}_0 \otimes \mathbf{s}_0)(\mathbf{a}_0 \otimes \mathbf{s}_0)^T] \nonumber\\
    &= \EX_X[\mb{a}_0\mb{a}_0^T \otimes \mb{s}_0\mb{s}_0^T]
\end{align}
and finally assuming that the expectation of Kronecker product is approximately equal to the Kronecker product of the expectations, which has no name so we simply call the \gls{kfac} approximation here, though technically more accurately is independence of the covariances of the activations and covariances of the sensitivities, 
\begin{align}
    F^{(l)} &\stackrel{\text{KFAC}}{\approx} \EX_X[\mb{a}_l^T\mb{a}_l] \otimes \EX_X[\mb{s}_l^T\mb{s}_l] \label{eq:kfac_layer} \\
    &= A_l \otimes S_l.
\end{align}
In practice, it is common to smooth the approximation to the Fisher in time. At any iteration in the optimization the quality of the approximation is limited by the number of samples in a batch. This can be unstable. Instead, the approximation is smoothed by replacing the left and right Fisher factors with their moving averages. There are different methods for implementing this exponentially decaying average of the history, in this work it is computed 
\begin{align}
    \bar{A}_{lt} = (\bar{A}_{l(t-1)} + \kappa A_{lt}) / \vkappa_t \label{eq:kfac_expdecay}
\end{align}
where $\vkappa_t = \vkappa_{(t-1)} + \kappa$ is defined as the total weight and $\vkappa_0 = 0$, $\bar{A}_{l0}$ is a matrix of zeros and $A_{lt}$ is the instantaneous covariances computed from Equation~\eqref{eq:kfac_layer}. The right Fisher factor is computed in the same way. It is important to be careful here that the contribution to the exponentially decaying average of the moving covariances from the zeroth term is zero, otherwise the following computed averages will be zero biased. The first update step is at $t=1$. 

The approximate natural gradients of each Fisher block can be computed as Equation~\eqref{eq:natural_gradients},
\begin{equation}
    \tilde{\delta}_l = \tilde{F}^{(l-1)} \mathrm{vec}(\Delta\mathcal{L}_l) \label{eq:kfac_gradients}
\end{equation}
We use $\tilde{\delta}$ to indicate approximate natural gradients throughout. 

\subsection{Damping}

This form of \gls{kfac} is unstable for two reasons: The approximate natural gradients can be large and the Fisher blocks are potentially singular (thus the inversion can't be performed). 

Damping is a tool for shifting the eigenspectrum of a matrix (such that the matrix is non-singular) and also restricting the `trust region' of the updates. The inverse Fisher described in Equation~\eqref{eq:kfac_gradients} is adjusted, typically in Tikhonov damping,
\begin{equation}
    \bar{\Delta}_l = (\tilde{F}^{(l)} + \lambda)^{-1} \mathrm{vec}(\Delta\mathcal{L}_l)
\end{equation}
This can no longer be decomposed as Equation ~\ref{eq:kronecker_inverse_identity}. An adapted technique, Factored Tikhonov Damping described in detail in Reference~\cite{martens2015optimizing}, is used which approximates the true damping
\begin{equation}
    (\tilde{F}^{(l)} + \lambda)^{-1} \stackrel{\text{FT}}{\approx} (A + \sqrt{\pi\lambda}) \otimes (S + \sqrt{\lambda / \pi}) 
\end{equation}
where
\begin{equation}
    \pi_l = \frac{\Tr[A_l]\dim(S_l)}{\Tr[S_l]\dim(A_l)}
    \label{eq:factoredtikhonov}
\end{equation}
This method is derived from minimizing the residual from the expansion of damped Fisher block into Kronecker factors.  

\subsection{Centering}

All approximations to the Fisher block have thus far been computed via the derivatives of the log-likelihood of the normalized distribution $p(X)$ with respect to the weights. However, we compute $\mathcal{O} = \frac{d \log |\psi(X)|}{d \theta}$ which are the derivatives of the log unnormalized wave function $\psi(X)$. It can be shown, for example Appendix C of Reference~\cite{pfau2019ab} that elements in the \gls{fim} can be expressed 
\begin{align}
    F_{ij} \propto \EX_X\Big[ (\mathcal{O}_i - \EX_X[\mathcal{O}_i]) (\mathcal{O}_j - \EX_X[\mathcal{O}_j]) \Big] \\
    = \frac{1}{4} \EX_X\bigg[ \frac{d \log p(X)}{d \theta_i} \frac{d \log p(X)}{d \theta_j} \bigg]
\end{align}
Where the derivatives of the log wave function have been centered by $\EX_X[\mathcal{O}]$. However, in tests we found that the performance was more dependent on the hyperparameters used in \gls{kfac} than on the centering. There are many approximations affecting the quality of the approximation to the \gls{fim} and the interplay between the optimization and this approximation. We ignored the centering to simplify the implementation but outline the method here of approximating this centering for readers who are interested.
\begin{align}
    F &\propto \EX_X\Big[ (\mathcal{O} - \EX_X[\mathcal{O}])^T (\mathcal{O} - \EX_X[\mathcal{O}]) \Big]  \nonumber \\
    &= \EX_X\Big[ \mathcal{O}^T\mathcal{O} - \EX_X[\mathcal{O}]^T \mathcal{O} - \mathcal{O}^T\EX_X[\mathcal{O}] \nonumber \\ & - \EX_X[\mathcal{O}]^T\EX_X[\mathcal{O}] \Big] \nonumber \\
    &\stackrel{\text{LINEAR}}{=} \EX_X[ \mathcal{O}^T\mathcal{O}] - \EX_X[\mathcal{O}]^T\EX_X[ \mathcal{O}] \label{eq:centering} \\
\end{align}
The second term can be approximated by assuming \gls{iad}:
\begin{align}
    \EX_X[\mathcal{O}]^T\EX_X[ \mathcal{O}] &= \EX_X[\mb{a} \otimes \mb{s}]^T\EX_X[\mb{a} \otimes \mb{s}] \\
    &\stackrel{\text{IAD}}{\approx} (\EX_X[\mb{a}] \otimes \EX_X[\mb{s}])^T (\EX_X[\mb{a}] \otimes \EX_X[\mb{s}]) \\
    &= \EX_X[\mb{a}]^T \EX_X[\mb{a}] \otimes \EX_X[\mb{s}]^T\EX_X[\mb{s}] \\
    &= A' \otimes S'.
\end{align}
Overall, the centering outlined in Equation~\eqref{eq:centering} can be approximated by mean centering the activations and sensitivities in Equation~\eqref{eq:kfac_layer}
\begin{align}
    \bar{\mb{a}} &= \mb{a} - \EX_X[\mb{a}] \\
    \bar{\mb{s}} &= \mb{a} - \EX_X[\mb{s}]
\end{align}
or alternately maintaining a moving average of $A'$ and $S'$ and subtracting those directly from $\bar{A}$ and $\bar{S}$ in Equation~\eqref{eq:kfac_expdecay}. Both these methods result in the same residual, which can be computed by expanding either expression for the resultant approximate \gls{fim}.


\section{Variational Monte Carlo}

Table~\ref{tab:kfacvariables} contains all the hyperparameters, variables and initial values for both the \gls{vmc} and \gls{kfac} methods used in this work. The entire algorithm, the update loop for the model, is given in Algorithm~\ref{alg:vmckfac}.

Figure~\ref{fig:fn_vmc_finalz} shows the \gls{vmc} evolution of the Ansatz on the systems Be - Ne. The evolution is roughly the same as previous work. Here, no explicit comparison is made to alternate optimizers. We note that in tests we found similar results to the original work where \gls{kfac} outperformed another method (ADAM).

\begin{table}[H]
\centering
\begin{tabular}{|l|l|l|}
\hline
Name                   & symbol        & initial value \\
\hline
iteration & $k$ & 0 \\ 
number iterations & $k^\text{VMC}_{\text{max}}$ & $1 \times 10^5$  \\
learning rate         & $\eta$ & $1 \times 10^{-4}$  \\
norm constraint  & $c$ & $1 \times 10^{-4}$ \\
covariance state decay   & $\kappa$ & 0.95  \\
damping  & $\lambda$ & $1 \times 10^{-4}$ \\
damping factor & $\pi$  & N/A \\
instantaneous activations  & $\mb{a}$ & N/A \\
instantaneous sensitivities  & $\mb{s}$ & N/A\\
energy gradients & $\Delta\mathcal{L}$  & N/A\\
approx. natural gradients  & $\tilde{\delta}$ & N/A\\

variable decay  & $\nu$ & $1 \times 10^{-4}$ \\
spatial locations  & $i$ & N/A\\

decaying average left Kronecker factor  & $ \bar{A}_l$ & $\mb{0}$ \\
decaying average right Kronecker factor & $ \bar{S}_l$ & $\mb{0}$\\
instantaneous left Kronecker factor & $A_l$ & N/A \\
instantaneous right Kronecker factor & $ S_l$  & N/A \\
correlation length & $c_l$ & 10 \\
sampling step size & $\sigma$ & $0.02^2$ \\
batch size & $m$ & 4096 \\
\hline

\end{tabular}
\caption{Table containing all variables used in the \gls{vmc} algorithm using \gls{kfac} updates in this work. \label{tab:kfacvariables}}
\end{table}

\begin{figure*}
\begin{minipage}{\linewidth}
\begin{algorithm}[H]
\caption{\gls{kfac} for Fermi Net*. \textbf{CholeskyInverse} indicates the Cholesky Inversion, \textbf{FactoredTikhonov} is the technique for computing $\pi$ given by Equation~\eqref{eq:factoredtikhonov}, and \textbf{Clip} indicates clipping a batch of values to within 5x of the median value. A high level overview of the distribution of this algorithm across multiple workers is shown in Figure~\ref{fig:kfac_distribution} \label{alg:vmckfac}}
\begin{algorithmic}[1]
\For{$k^{\text{VMC}}_\text{max}$}
    \item[]
    \State Update walker coordinates $X$ \algorithmiccomment{Metropolis Hastings}
    \State $\tilde{E}_L = \textbf{Clip}(E_L(X) - \EX_X[E_L(X)])$ \algorithmiccomment{energy}
    \State $\Delta \mathcal{L} = \EX_X \Big[\tilde{E}_L \mb{\nabla} \log|\psi(X)|  \Big]$ \algorithmiccomment{ backward pass}
    \State Compute $\mb{a}$ \algorithmiccomment{forward pass}
    \State Compute $\mb{s}$ \algorithmiccomment{backward pass}
    \item[]
    
    \For{all layers $l$}   
        \item[]

        \State $\bar{\mb{a}}_l \gets \EX_i[\mb{a}_{li}]$
        \State $A_l \gets \EX_X[\bar{\mb{a}}_l^T\bar{\mb{a}}_l]$
        \State $\bar{A}_l \gets \kappa \bar{A}_l + (1 - \kappa) A_l$
        \item[]
        
        \State $\bar{\mb{s}}_l \gets \EX_i[\mb{s}_{li}]$
        \State $S_l \gets \EX_X[\bar{\mb{s}}_l^T\bar{\mb{s}}_l]$
        \State $\bar{S}_l \gets \kappa \bar{S}_l + (1 - \kappa) S_l$
        \item[]
        
        \State $\pi_l \gets \textbf{FactoredTikhonov}(\bar{A}_l, \bar{S}_l)$
        \State $\lambda_{A} \gets (\frac{\lambda}{\pi \times |T|^2})^{1/2}$
        \State $\lambda_{S} \gets (\frac{\lambda \times \pi}{|T|^2})^{1/2}$
        \item[]
        
        \State $\bar{A}_l^{-1} \gets $\textbf{CholeskyInverse}$(\bar{A}_l + I\lambda_A)$
        \State $\bar{S}_l^{-1} \gets $\textbf{CholeskyInverse}$(\bar{S}_l + I\lambda_S)$
    
        \item[]
        
        \State $\tilde{\delta}_l \gets \bar{A}_l^{-1} \frac{\Delta_l \mathcal{L}}{|T|^2} \bar{S}_l^{-1}$ 
        \item[]
    \EndFor
    \item[]
    \State $\epsilon \gets \min\Big( 1, \sqrt{\frac{c}{\sum_l \tilde{\delta}_l \Delta_l \mathcal{L}}} \Big)$
    \item[]
    
    \For{all layers $l$}
        \item[]
        \State $\theta_l \gets \theta_l - \epsilon \times \eta \times \tilde{\delta}_l$ \algorithmiccomment{Update the model}
        \item[]
    \EndFor
    \item[]
    
    \State $\eta \gets \frac{\eta_0/\nu}{1 +  k}$
    \State $\lambda \gets \frac{\lambda_0/\nu}{1 +  k}$
    \State $c \gets \frac{c_0/\nu}{1 +  k}$
    \item[]
\EndFor
\end{algorithmic}
\end{algorithm}
\end{minipage}
\end{figure*}

\section{Diffusion Monte Carlo}\label{app:dmc}

Table~\ref{tab:dmc_variables} contains all the hyperparameters, variables and initial values used in the version of \gls{dmc} implemented for this work. The entire algorithm is outlined in Algorithm~\ref{alg:dmc}.
\begin{table}[H]
\begin{tabular}{|l|l|l|}
\hline
Name                   & symbol        & initial value                \\ \hline
number iterations       & $k^{\text{DMC}}_{\text{max}}$ & variable \\
Forces                  & $\mb{F}$ & N/A \\
Green's function        & $G(\mb{r}, \mb{r}')$ & 1. \\
trial energy  &  $E_T$  & N/A \\
acceptance (rejection) probability    & p (q)     & N/A \\
 weights & $\boldsymbol{\omega}$   & $\mb{1}$ \\
batch size & $M$ & 4096 \\
    &    &  \\
\hline

\end{tabular}
\caption{Variables used in the Diffusion Monte Carlo method. \label{tab:dmc_variables}}
\end{table}
\begin{figure*}
\begin{minipage}{\linewidth}
\begin{algorithm}[H]
\caption{Diffusion Monte Carlo. $\mb{r}$, $\boldsymbol{\xi}$ and $\mb{F}$ are $M \times n_e \times 3$ dimensional tensors. Steps update electron positions simultaneously and acceptance of steps are performed in parallel across all walkers. Line 7 updates walkers where the condition is true. \label{alg:dmc}}
\begin{algorithmic}[1]
\item[]
\State Compute $\psi(X)$  \algorithmiccomment{forward pass}
\State $\mb{F} \gets \frac{d\log|\psi(X)|}{d\mb{r}}$ \algorithmiccomment{backward pass}
\item[]
\For{$k^\text{DMC}_\text{max}$}
    \item[]
    \State $G(\mb{r}, \mb{r}') = G(\mb{r}', \mb{r}) = 1$
    \item[]
    
    \State $\boldsymbol{\xi} \sim N(0, \tau)$
    
    \State $\mb{r}' \gets \mb{r} + \tau \mb{F} + \boldsymbol{\xi}$
    
    \item[]
    \State $X' \gets \{\mb{R}, \mb{r}'\}$
    \State Compute $\psi(X')$ \algorithmiccomment{forward pass}
    \State $\mb{F}' \gets \frac{d\log|\psi(X')|}{d\mb{r}'}$ \algorithmiccomment{backward pass}
    \item[]
    
    \State $G(\mb{r}, \mb{r}') \gets \prod_{i}^{n_e} \exp\big((\mb{r}_i - \mb{r}_i' - \tau \mb{F}_i')^2 / 2\tau  \big)$
    \State $G(\mb{r}', \mb{r}) \gets \prod_{i}^{n_e} \exp\big((\mb{r}'_i - \mb{r}_i - \tau \mb{F}_i)^2 / 2\tau  \big)$
    
    \item[]
    \State $p \gets \min\Big(1, \frac{|\psi(X')|^2 G(\mb{r}, \mb{r}')}{|\psi(X)|^2 G(\mb{r}', \mb{r})}\Big)$
    \State $q \gets 1 - p$
    
    \State Set $p = 0$ where $\mathrm{sign}(\psi(X)) \neq \mathrm{sign}(\psi(X'))$ \algorithmiccomment{Fixed node approximation}
    \item[]
    
    \State $s \gets E_T - E_L(X)$
    \State $s' \gets E_T - E_L(X')$
    
    \State $\omega \gets \omega \times \exp\Bigg[\tau\Big[\frac{p}{2}\big(s' + s \big) + qs  \Big]\Bigg]$
    \State $\alpha \sim U[0, 1]$
    \If{$p > \alpha$}
        \State $\psi(X) \gets \psi(X')$, $\mb{r} \gets \mb{r}'$, $\mb{F} \gets \mb{F}'$, etc \algorithmiccomment{Update variables for next iteration}
    \EndIf
\EndFor
\end{algorithmic}
\end{algorithm}
\end{minipage}
\end{figure*}

\clearpage

\bibliography{bibliography}

\end{document}